# Stick-slip Phenomena and Memory Effects in Moving Vortex Matter

*Lise Serrier-Garcia[1*], Clécio C. de Souza Silva[2], Matias Timmermans[1], Joris Van de Vondel[1] and Victor V. Moshchalkov[1]*

[1] INPAC – Institute for Nanoscale Physics and Chemistry, KU Leuven, Celestijnenlaan 200D, B–3001 Leuven, Belgium

[2] Departamento de Fisica, Universidade Federal de Pernambuco, Cidade Universitaria, 50670-901 Recife, Pernambuco, Brazil.

* e-mail: serriergarcia.lise@fys.kuleuven.be

**Manipulating vortices in non-conventional superconductors is nowadays a challenging path toward controlling functionalities for superconducting nanodevices [Grigorenko 2001, Auslaender 2008, Cordoba 2013, Guillamon 2014, Roditchev 2015, Embon 2015, Zhou 2015]. Here, we directly observe and control single vortex core trajectories with unmatched resolution using a new scanning tunneling spectroscopy at very low temperature. Our data show the depinning threshold of a Bragg-glass in a weakly disordered superconductor, a clean 2H-NbSe$_2$ crystal. We first experimentally capture the linear and collective response, the Campbell regime [Campbell 1972]. Upon strong drives, the oscillating trajectories perform a series of stick-slip motions that mimics the lattice periodicity. We then theoretically elucidate this peculiar non-linear regime by solving the Langevin dynamics equations. We additionally explore the impact of initial conditions and reveal an**

**enhancement of the long-range correlations with the cooling procedure [Larkin 1979]. Finally, our work establishes a connection between theory of vortex pinning, memory effects and vortex lineshapes, thus offering a new platform to investigate the relationship between viscous media and individual controllable objects in any many-body systems.**

Vortices in type-II superconductors are quantized whirlpools of supercurrents which repel each other and form an approximately periodic medium dubbed vortex matter. Like other systems of interacting bodies, such as polymers [Doi 1986], Fermi gases [Zwierlein 2005], glass-forming liquids [Dyre 2006] and plasma crystals [Schweigert 1998], vortices are sensible to thermal fluctuations as well as to disorder in the host environment. In a superconducting crystal, the disorder is provided by intrinsic defects which pin the vortex lines [Blatter1994]. The ease of controlling the vortex density, defect concentration and temperature makes vortices a very attractive system in which to study the rich variety of phenomena stemming from the interplay between elastic, pinning and thermal forces. Indeed, the recent development of powerful experimental tools, especially high resolution imaging techniques, spawned important advances in this area [Auslaender 2008, Embon 2015].

The idea of correlated vortex bundles, also known as Larkin domains, was introduced in the framework of the collective pinning theory [Larkin1979] and further developed by more involved theories [Feigel'man1989, Giamarchi1994]. Although this concept has been useful to estimate important static properties of superconductors, such as the maximum non-dissipative current carried by the material, it is still unclear how a Larkin domain evolves under the application of a driving force. It is generally believed that the onset of vortex motion in weakly disordered superconductors coincides with the plastic breakdown of the Larkin domains [Moon 1996]. At higher drives, this plastic regime gives place to a new elastic phase, called moving Bragg glass, in which vortices again arrange into correlated volumes, but this time adjusted to a dynamic pinning potential [Pardo 1998]. However, for very weak pinning, it has been predicted that the plastic phase is suppressed and the vortex bundles depin while keeping their integrity [LeDoussal 1998]. This situation opens the unique opportunity of studying the *dynamic properties* of the Larkin domains across the depinning threshold. Previous imaging experiments have identified coherent dynamics of vortices in pristine $NbSe_2$ single crystals at long time scales, in which case the dynamics is dominated by slow thermal relaxation processes [Troyanovski 1999, Lee 2011]. Here, using a new synchronized scanning tunneling spectroscopy technique (see Methods and SI2), we are able to resolve much faster processes occurring in a clean $NbSe_2$ at a very low temperature (0.5 K), where thermal fluctuations can be neglected. This allows us to investigate in detail vortex trajectories resulting from the interaction between the vortex bundle and the underlying effective potential.

We have started dealing with the complexity of the phase diagram by probing the starting motion until the depinning threshold of a rigid vortex lattice inside a Larkin domain. We evidence that at low drive it reversible oscillating motions follow the alternative driving force in Fig. 1a-c, while at high drives the trajectories present striking nonlinear trajectories with a pinning-depinning process instead of the expected free motion (see in Fig 2a-d). Performing three-dimensional Langevin dynamics simulations, we demonstrate in Fig. 3a-c that the jerky motion is governed by the translation of the vortex lattice to reach a unique metastable state. Finally, history-dependent signatures in space and time of the trajectories provides in Fig. 4a-b creeps at huge timescales and anisotropy of currents in the c-direction, oppening possible clues to explain the memory effects observed in high-Tc superconductors.

At low drive, each vortex in Fig. 1b-c, vibrates in the free space between two first neighbors that, if considering that the vortex lattice is built up by single objects connected by springs, may correspond to the soft elastic modes of the vortex lattice as suggested in ref. [Valenzuela 2002] and consistently with ref [Timmermans 2014]. The trajectories in Fig. 1b-c indicate that locally the vortex lattice oscillates within a single potential energy well and is restored by the local force, $F_{res} = -\alpha u$, for a displacement $u$, where $\alpha$ is the Labush constant determining the curvature of the harmonic potential well [Campbell 1972]. The vortex traces are well-defined, indicating a reversible and deterministic motion at 0.5 K (SI4). This primary result showing that individual trajectories cycle with the oscillating external excitation is the first direct and local observation of the *linear response* of a superconducting condensate.

A large data set reveals that the linear trajectories are limited to low drives. The dynamic consists at stronger drive in elongated ellipses deviated toward the vortex lattice axis direction. The vortex lattice still coherent and oscillating, hopes between the two extrema of the trajectory, reaching high velocities up to 140 μm/s well above the 10 μm/s of the linear regime (Fig. 2c). A thorough investigation of the nonlinear regime indicates that the trajectories have four important features: (i) the number of observed attractive points (noted A, B, … ) is exactly two or three (SI7-8), (ii) the distance between these points increases step by step with the drive (see inset in Fig. 3c), (iii) the hopping between points occurs at a single value of the drive, $H_{ac}(t_{jump})$, independent of the cooling procedure (Fig. 2c), (iv) trajectories obtained with the same drive procedure are reproducible, although they are randomly shifted in the ab-plane.

Following the idea that the nonlinear regime is the consequence of a collective effect, we performed Langevin dynamics simulations of a three-fold lattice composed by flexible line subjected to a random pinning potential (see methods and SI3). For low amplitudes, the system oscillates within a single potential energy well, the linear regime. Above a threshold value, the vortices jump coherently between two different low energy positions, revealed in the time dependent of the mean vortex speed by one strong peak at each branch of the excitation cycle, similar to the experimental nonlinear regime (blue and red curves in Fig. 3b). As the amplitude of the drive is further increased, new steps in the mean excursion range regularly appear, in conjunction with the creation of vortex lattice hopping, between exactly $n$ wells at the $n$-eenth step (Fig. 3a-c and SI11-12). More than strikingly imitating the experimental step by step regimes, the simulations evidence an *excursion-lock motion*.

At this step, although the jerky motion is reproduced, the necessity of a pinning landscape is not clear and we thus calculated separately the time-averaged pinning, line and inter-vortex components of the total free energy. It appears that because of their strong mutual interaction, vortices keep their lattice structure and thereby the mean inter-vortex and line energy fluctuate slightly about constant values (SI10). The dynamics is therefore dominated by the pinning potential probed by the vortex bundle. Hence, the driving force dependence of the pinning energy presents a series of minima at odd steps (1, 3 and 5 in Fig.4c) which remarkably corresponds to integer multiples of the vortex lattice spacing (respectively 0, 1 and 2 times $a_0$). In opposition to the random nature of the local pinning potential, the effective pinning landscape inherits the periodicity of the moving vortex system as a result of its *discrete translational invariance*. Since translations coinciding with lattice vectors make the system return to its more favorable minimum energy configuration, such modes are natural attractors of the dynamics. As an aside, an extra motion is visible at each extremity of the trajectory, reflecting the local range $\xi_v$ of the pinning energy, in other terms, the minimum size of the potential well. The stick-slip motion thus appears related to the correlations of the lattice in addition to its elastic properties.

Several study based on the resistivity measurement provides that the range of lateral order may enlarge with annealing the vortex lattice by a drive [Yaron 1994, Paltiel 2000, Valenzuela 2000, Pérez Daroca 2010, Raes 2014]. In order to prove this assumption, we applied different ac-modulations during cooling (Fig. 2a-d). We observed an enhancement of the motion amplitudes in both linear and nonlinear regimes, suggesting a reduction of the local Labush parameter. Hence, when the increased excitation is tuned on, the system *remembers* the quenched state and presents an enhanced response to the excitation field. As such, the stick-slip regime is robust against various preparation conditions and only the effective local drive is reinforced, certainly related to enlarge Larkin domains.

Assuming that the local driving force is caused by current flow and vortex compression in the ab plan, one can expect it to be inhomogeneous approaching the surface sample along the c-axis, causing dynamics frustration in the c-direction. We believe that anisotropy in the currents or pinning layers along the c-axis provides a lower drive of the vortex lattice sections deeper into the material. The later consequently appeal the free vortex segments of the top layer to follow their path via the vortex line tension, causing a robust *dynamical memory effect* in space and time. Fig. 4c illustrates this behavior. The motion relaxed collectively from the first to the second regime after increasing the shaking (SI9). The nonlinear trajectory reproduces the previous path still followed by deeper vortex segments and appears consequently asymmetric (see other examples in SI5 and SI7-8).

Whereas the nonlinearities in vortex dynamics are sensitive to frequency, we observed that motions are independent of low frequencies (15 -122 Hz in SI5). This confirms the unimportance of both thermal fluctuations and viscosity at the millisecond timescale in our case. However, we observed *relaxations* of the system in much larger timescales, by comparing trajectories obtained at 77 Hz with those performed in the $10^{-4}$ Hz range (Fig. 3a). In fact, the time needed to reach equilibrium positions by means of thermal or quantum fluctuations leads to a creep motion of vortices in a weakly-rough potential [Anderson 1962, Campbell 1972]. The constant driving force pushes vortices to slide at each step of the mHz-case, while the instantaneous drive of the Hz-case is reduced before the end of this relaxation process. The later trajectories are consequently a miniature of the former. One can note the good agreement with creep timescales of truly three-dimensional vortex-glass phases [Thompson 1991, Gordeev 1997, Pardo 1998, Henderson 1998], leading us to think that this low relaxation deals more generally with disordered vortex matter in high-Tc superconductors such as cuprates [Valenzuela 2000, Grigorenko 2001] or pnictides [Moll 2014]. Furthermore, we confirmed the *roughness* of the potential by suddenly stopping the driving force during the ac-cycle in a second experiment (Fig. 3b). Comparing square and circular symbols at each time $t_i$ shows that the vortex lattice is quenched at the present location without relaxation. We explain the absence of relaxation in this second case by the huge timescale of the creep motion at low temperature in absence of an external drive.

In conclusion, we directly visualize individual trajectories with unmatched resolution while experimental results were almost limited to indirect techniques integrating a large amount of vortices or local probes averaging fast motions in time. Subsequently, we first unveil the linear response a Bragg vortex glass cycling with the oscillating drive and then solve the ununderstood stick-slip paths already observed in $NbSe_2$ but never explained. The success of this work opens new possibilities to solve issues such as the dynamical channels related to the wave density charge in $Nbse_2$ or the memory effects in high-Tc superconductors. Finally, we conclude that these effects revealed in superconductors are generic to any elastic periodic lattices in viscous media, they should manifest in a wide variety of many-body systems and the potential of our technique is powerful to investigate single-object externally controllable in soft matter.

## Methods

All the experiments were performed on a cleaved 2H-$NbSe_2$ with dimensions L x P x H ~ 2 x 2 x 0.5 mm (see SI1). The scanning tunneling spectroscopy measurements were performed with an Attocube apparatus, in a vacuum with P ~ $10^{-6}$ mbar and at a base temperature of 350 mK, while the electron temperature was estimated to be 600 mK. A mechanically sharpened Au tip was covered in situ by Pb, resulting in a superconducting tip. Typical set-point parameters for spectroscopy were 80 pA for 4 mV applied to the sample with respect to the tip. A constant magnetic field of 100 mT modulated by by $\vec{H}_{ac}(t) = H_{ac} \sin(2\pi t f)\, \vec{c}$ is applied perpendicular to the sample surface. An extra superconducting coil fitted around the STM head inside the vacuum shield provides the additional ac or stepping magnetic field in a maximum range of +/- 0.12 mT. Synchronized to this ac-modulation, the tunneling current is acquired at 1 mV at each spatial point in loop-off [Timmermans 2014], leading to 520 spatial maps per period (See SI2). The real-time trajectories are numerically fitted from these data and quasi-instantaneous velocities are numerically derivate from the time-trajectory.

The simulations were performed using Langevin-Bardeen-Stephen equations in a 1.20 x 1.17 x 2.40 µm² piece of a superconducting crystal. The material parameters were chosen similar to typical values of $NbSe_2$ (SI3). The system comprise $N_v = 72$ vortex lines with periodic boundaries conditions in the *ab* planes, thus corresponding to a vortex state lattice spacing $a_0 = 150\,nm$. As the simulated trajectories follow the driving direction, we arbitrary select direction parallel to a principal axis of the triangular vortex lattice. The driving force is induced by superconducting currents homogeneously-distributed in the simulation box, and follows the sinusoidal excitation. Since the frequency of 77 Hz induces huge computing time, we have chosen the 0.4 $MHz^{-1}$ well above the Ginzburg-Landau time of a superconducting condensate ($t_{GL} \cong 10^{12} s$). The root-mean-square of the pining potential is $U_0 = 81\,K$, well above the thermal enegy available at the temperature of 0.5 K, ensuring a creep time much longer than several cycles of the excitation force (see Fig. 4). The pining potential is modeled by a dense random distribution of pinning energies smoothed over a length scale equal to the vortex core radius $\xi_v = \sqrt{2}\xi$ [Blatter 1994].

**References**

Grigorenko, A., Bending, S., Tamegai, T., Ooi, S. Henini, M., A one-dimensional chain state of vortex matter, *Nature* **414** 728-731 (2001)

Auslaender, O. M., Luan, L., Straver, E. W; J., Hoffman, J. F., Koshnick, N. C., Zeldov, E., Bonn, D. A., Liang, R., Hardy, W.N., Moler, K. A., Mechanics of individual isolated vortices in a cuprate superconductor, Nat. Phys. 5 35-39 (2008)

Córdoba, R., Baturina, T. I., Sese, J., Mironov, A. Y., De Teresa, J. M., Ibarra, M. R., Nasimov, D. A., Gutakovskii, A. K., Latyshev, A. V., Guillamón, I., Suderow, H., Vieira, S.,

Baklanov, M. R., Palacios, J. J., Vinokur, V. M., Magnetic field-induced dissipation-free state in superconducting nanostructures, *Nat. Com.* **4** 1437 (2013)

Guillamón, I., Córdoba, R., Sesé, J., De Teresa, J. M., Ibarra, M. R., Vieira, S., Suderow, H., Enhancement of long-range correlations in a 2D vortex lattice by an incommensurate 1D disorder potential, *Nat. Phys.* **10** 851-856 (2014)

Roditchev, D., Brun, C., Serrier-Garcia, L., Cuevas, J. C., Loiola Bessa, V. H., Miloševic, M. V., Debontridder, F., Stolyarov, V., Cren, T., Direct observation of Josephson vortex cores, *Nat. Phys.* **11** 332-337 (2015)

Embon, L., Anahory, Y., Suhov, A., Halbertal, D., Cuppens, J., Yakovenko, A., Uri, A., Myasoedov, Y., Rappaport, M. L., Huber, M. E., Gurevich, A., Zeldov, E., Probing dynamics and pinning of single vortices in superconductors at nanometer scales, *Sci. Rep.* **5** 7598 (2015)

Zhou, C., Reichhardt, C., Olson Reichhardt, C. J., Beyerlein, I. J., Dynamic phases, pinning, and pattern formation for driven dislocation assemblies, *Sci. Rep.* **5** 8000 (2015)

Campbell, A. M., Evetts, J. E., Flux vortices and transport currents in type-II superconductors, *Adv. Phys.* **21** 377-577 (1972) *Adv. Phys*. **50** 1249-1449 (2001)

Larkin, A. I., Ovchinnikov, Y. N., Pinning in type-II Superconductors, *J. Low. Temp.* **34** 409-428 (1979)

Doi, M., Edwards, S. F., *The theory of polymer dynamics* (Clarendon 1986)

Zwierlein, M. W., Abo-Shaeer, J. R., Schirotzek, A., Schunck, C. H., Ketterle, W., Vortices and superfluidity in a strongly interacting Fermi gas, *Nature* **435**, 1047-1051(2005)

Dyre, J. C. The glass transition and elastic models of glass-forming liquids. *Rev. Mod. Phys*. **78**, 953–972 (2006)

Schweigert, V. A., Schweigert, I. V., Melzer, A., Homann, A., Piel, A., Plasma crystal melting a nonequilibrium phase transition, *Phys. Rev. Lett.* **80** 5345-5348 (1998)

Blatter, G., Feigel'Man, M. V., Geshkenbain, V. B., Larkin, A. I., Vinokur, V. M., Vortices in high-temperature superconductors, *Rev. Mod. Phys.* **66** 1125-1388 (1994)

Feigel'man, M. V., Geshkenbein, V. B., Larkin, A. I., Vinokur, V. M., Theory of collective flux creep, *Phys. Rev. Lett*. **63** 2303-2306 (1989)

Moon, K., Scalettar, R. T., Zimányi, G. T., Dynamical phases of driven vortex systems, *Phys. Rev. Lett.* **77** 2778-2781 (1996)

Pardo, F., de le Cruz, F., Gammel, P. L., Bucher, E., Bishop, D. J., Observation of smectic and moving-Bragg-glass phases in flowing vortex lattices, *Nature* **396** 348-350 (1998)

Le Doussal, P., Giamarchi, T., Moving glass theory of driven lattices with disorder, *Phys. Rev. B* **57** 11356-11403 (1998)

Troyanovski, A. M., Aarts, J., Kes, P. H., Collective and plastic vortex motion in superconductors at high flux densities, *Nature* **399** 665-668 (1999)

Lee, J., Wang, H., Dreyer, M., Berger, H., Barker, B. I., Nonuniform and coherent motion of superconducting vortices in the picometer-per-second regime, *Phys. Rev. B* **84** 060515(R) (2011)

Valenzuela, S. O., Order and mobility of solid vortex matter in oscillatory driving currents, *Phys. Rev. Lett.* **88** 247003 (2002)

Timmermans, M., Samuely, T., Raes, B., Van de Vondel, J., Moshchalkov, V. V., Dynamic visualization of nanoscale vortex orbits, *ACS Nano* **8** 2782-2787 (2014)

Yaron, U., Gammel, P. L., Huse, D. A., Kleiman, R. N., Oglesby, C. S., Busher, E., Batlogg, B., Bishop, D. J., Neutron diffraction studies of flowing and pinned magnetic flux lattices in 2H-$Nbse_2$, *Phys. Rev. Lett.* **73** 2748-2751 (1994)

Paltiel, Y., Zeldov, E., Myasoedov, Y. N., Shtrikman, H., Bhattacharya, S., Higgins, M. J., Xiao, Z. L., Andrei, E. Y., Gammel, P. L., Bishop, D. J., Dynamic instabilities and memory effects in vortex matter, *Nature* **403** 398-401 (2000)

Valenzuela, S. O., Bekeris, V., Plasticity and memory effects in the vortex solid phase of twinned $YBa_2Cu_3O_7$ single crystals, *Phys. Rev. Lett.* **84** 4200-4203 (2000)

Pérez Daroca, D., Lozano, G. S., Pasquini, G., Bekeris, V., Depinning and dynamics of ac driven vortex lattices in random media, *Phys. Rev. B* **81** 184520 (2010)

Raes, B., de Souza Silva, C. C., Silhanek, A. V., Cabral, L. R. E., Moshchalkov, V. V., Van de Vondel, J., Closer look at the low-frequency dynamics of vortex matter using scanning susceptibility microscopy, *Phys. Rev. B* **90** 134508 (2014)

Anderson, P. W., Theory of flux creep in hard superconductors, *Phys. Rev. Lett.* **9** 309 (1962)

Thompson, J. R., Sun Yang Ren, Holzberg, F., Long-term nonlogarithmic magnetic relaxation in single-crystal $YBa_2Cu_30_7$ superconductors, *Phys. Rev. B* **44** 458-461 (1991)

Gordeev, S. N., de Groot, P. A. J., Oussena, M., Volkozurb, A. V., Pinfold, S., Langan, R., Gagnon, R., Taillefer, L., Current-induced organization of vortex motion in type-II superconductors, *Nature* **385** 324-326 (1997)

Henderson, W., Andrei, E. Y., Higgins, M. J., Plastic motion of a vortex lattice driven by alternating current, *Phys. Rev. Lett.* **81** 2352-2355 (1998)

Sudbø, A., Brandt, E. H., Nonlocal elastic properties of flux-line lattices in anisotropic superconductors, Phys. Rev. B **43** 10482 (1991)

Moll, P. J. W., Zhu, X., Cheng, P., Wen, H.-H., Batlogg, B., Intrinsic Josephson junctions in the iron-based multi-band superconductors $(V_2Sr_4O_6)Fe_2As_2$, *Nat Phys.* **10** 644-647 (2014)

**Acknowledgements**

We acknowledge financial support from Mathusalem funding 08/05 by the Flemish Government, the Fund for Scientific Research-Flanders (FWO-Vlaanderen), and the MP1201 COST-action.

**Author Contributions**

The experiments were conceptualized and analyzed by L.S.-G.. They were performed by L.S.-G. and M.T.. The theoretical simulations were provided and analyzed by C.C.S.S.. The manuscript was written by L.S.-G and C.C.S.S. with comments and inputs from all authors.

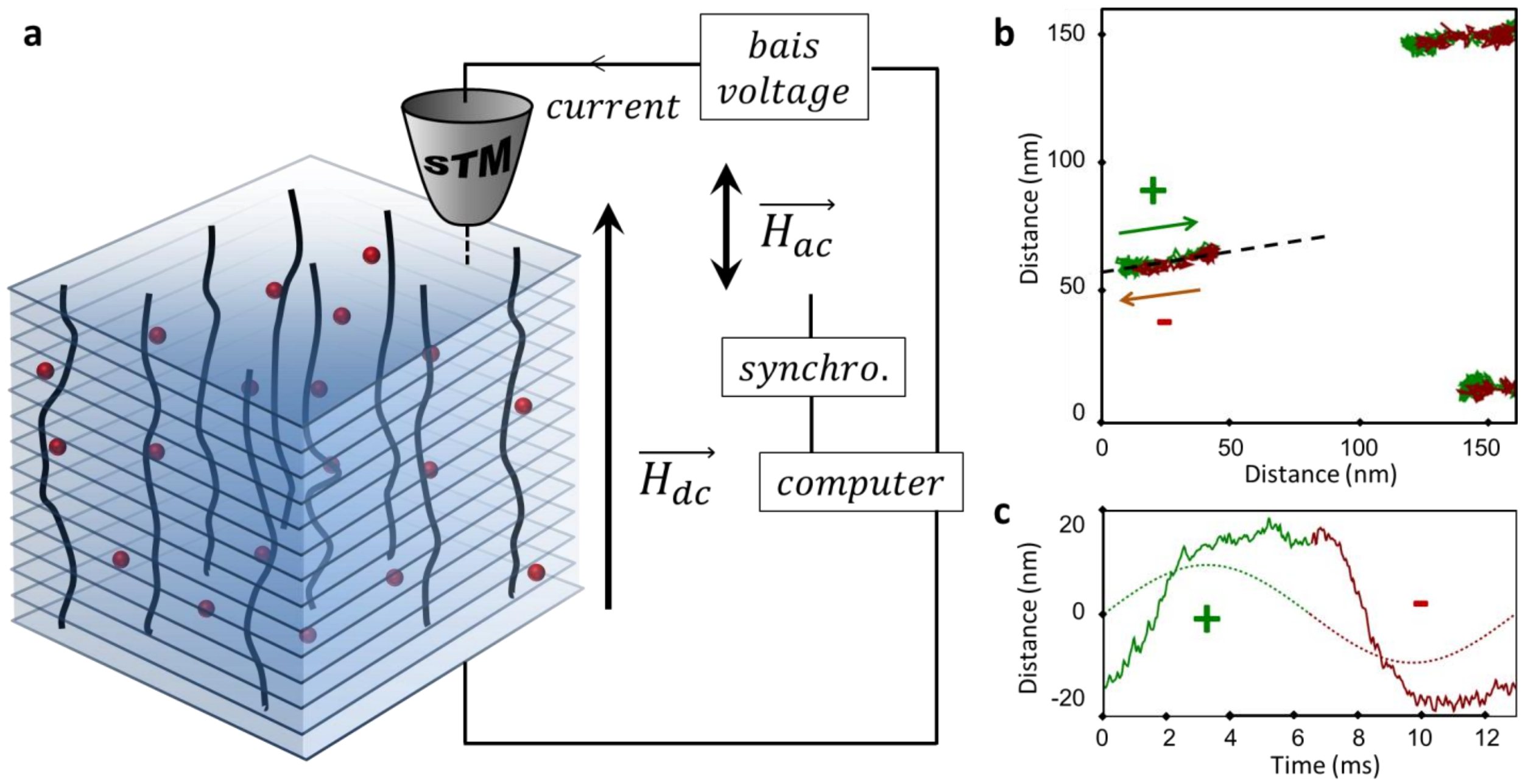

**Figure 1| The synchronized Lazy-Fisherman STM spectroscopy probing ac-driven vortices in $NbSe_2$ at 0.5 K.** A periodic motion is probed using an extended version of the "Lazy fisherman method" (SI2), in which a fisherman (STM tip) waits at a fixed position for a fish (vortex) to pass by. This method synchronized to the external excitation gives access to the *individual trajectory in real-time*. **a,** A dc-magnetic field is applied to generate superconducting vortices while a small modulation is superimposed to put them in motion. The STM measurement is synchronized to an external ac excitation, locally probing vortex trajectories in real-time. As dynamical effects depend on the interplay between the driven object (the vortex lattice) and its environment (the pinning landscape), the 3D properties of the layered superconductors $NbSe_2$ are investigated (See SI1-2). **b, c,** Trajectories of three neighboring vortices show collective motion under ac-drive. Each vortex vibrates in the direction between two first neighbors and reproduces the sinusoidal signal (dashed line) with a phase lag of $\pi/4$. (**b** 160 x 160 nm², $H_{ac|cool}/H_{ac|m}$@0/0.099mT)

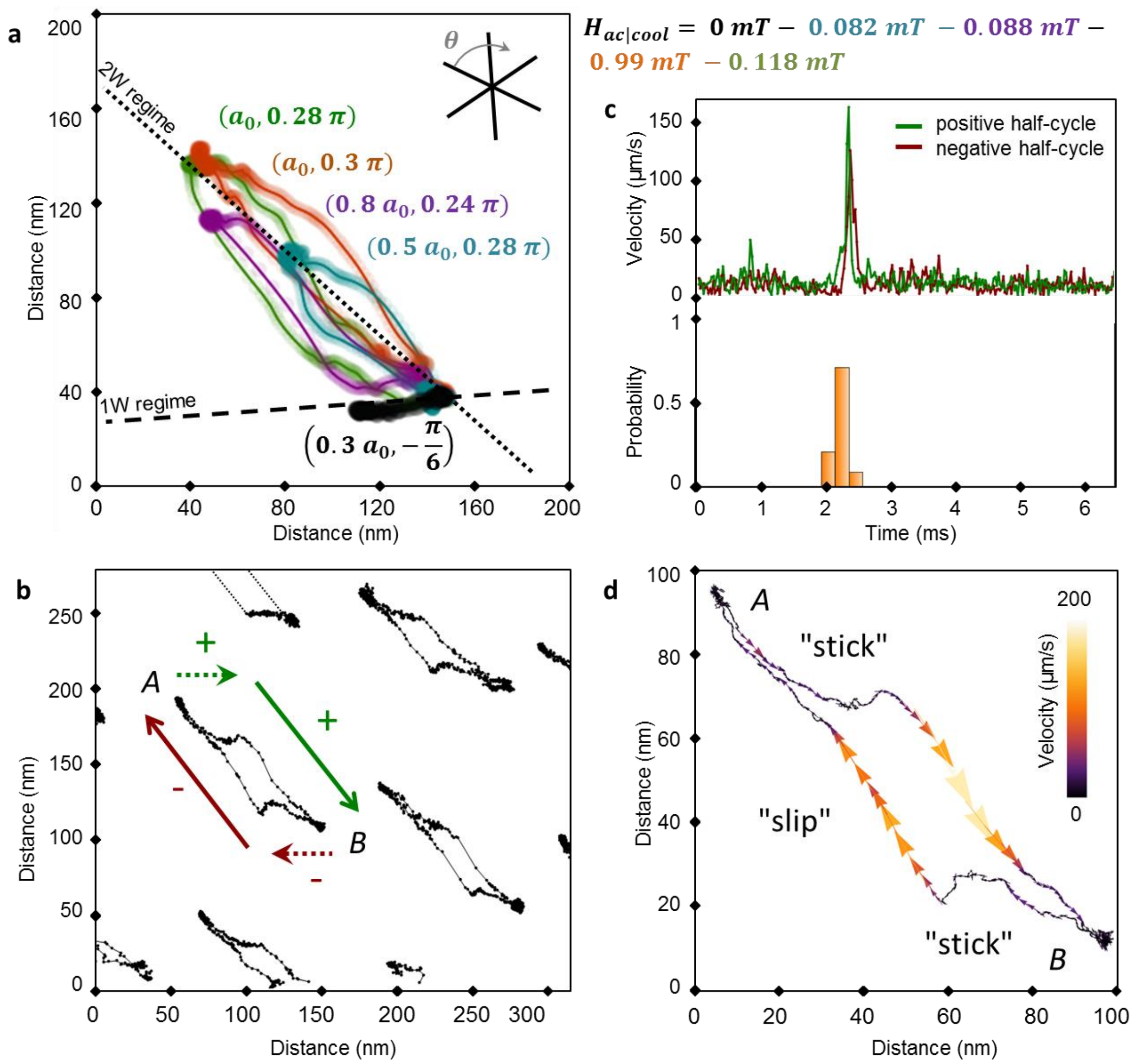

**Figure 2| Vortex trajectories vs amplitude of ac-excitation during preparation in $NbSe_2$ at 0.5 K. a-d,** The trajectory of the vortex lattice strongly depends on the applied ac-excitation during cooling. After quenching the system, a same shaking reveals two regimes: the linear trajectory (1W) presented in Fig. 1 and a nonlinear regime (2W). The former consists in a line in the -π/6 direction (dashed line) with low velocity (below 10 $\mu m.s^{-1}$), and the latter an ellipse along the 0.28 π direction (dotted line) with a high-velocity peak (up to 140 $\mu m.s^{-1}$). **a**, The

trajectories were measured at the same ac-excitation $H_{ac|m} = 0.099$ mT, after cooling the system at $H_{ac|cool} = 0$ (black), 0.082 (blue), 0.088 (violet), 0.099 (orange), 0.118 mT (green). As the position of the vortex lattice are different after each preparation, trajectories are shifted to match at t = 0. Orientation θ is referred to the vortex lattice (indicated by the black star) and the amplitude D in unit of the vortex spacing $a_0 = 135\ nm$. An error bar is indicated as a lighter contour around each curve (see SI3). **b**, The trajectories are measured during a large amount of ac-cycles, showing coherent and reproducible vortex motion. However, trajectories (a) and velocities (c-d) do not reproduce the sinusoidal excitation, and the vortex lattice stays mostly in points A and B. **c,** The velocity graphs during the positive (red) and negative (green) ac-modulation show a high-velocity peak in the nonlinear regime absent in the linear regime. The probability graph obtained for fifteen experiments demonstrates a unique value $t_{jump} = 2.19 \pm 0.05\ ms$. (**a** 200 x 200 nm², **b** 280 x 315 nm², **d** 100 x 100 nm² (zoom of **b**), **b** $H_{ac|cool}/H_{ac|m}$@0.099/0.099 mT).

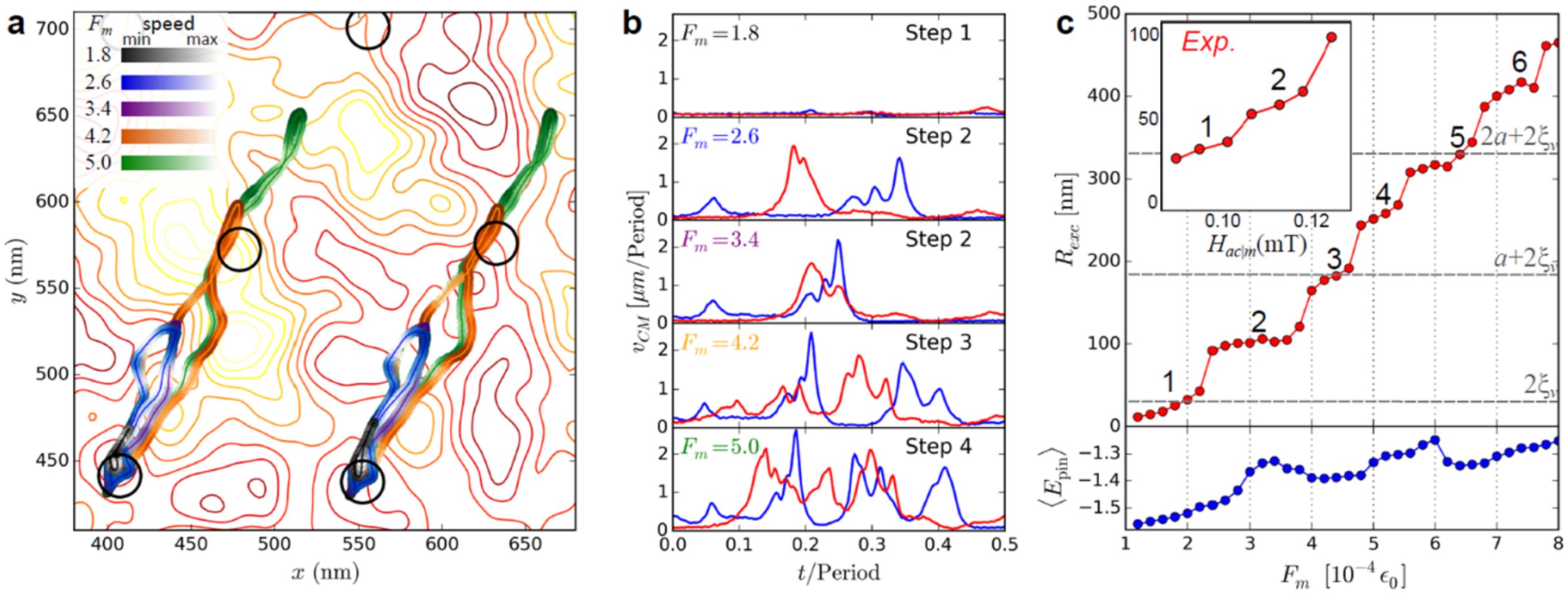

**Figure 3| Three-dimensional Langevin-Bardeen-Stephen simulations evidence a stick-slip motion of the rigid vortex lattice.** The dynamics are represented with stepping up the ac excitation amplitudes as following: trajectory in **a,** velocity in **b**, total excursion, line and pin energy in **c**. **a,** The motion of the top-layer elements of two neighboring vortices appears coherent and increase with $F_m$ for 1.8 (black), 2.6 (bleu), 3.4 (purple), 4.2 (orange) and 5.0 (green) (in units of $10^{-4}\epsilon_0 = 10^{-3}pN$). Similarity in trajectories of different $F_m$, whose the initial position at $F_m = 0$ (black circles), provide the low energetic vortex states. Because of the relatively strong vortex-vortex interactions and of the elastic coupling of the top vortex element to the rest of the line, the low velocity points (dark intensity) do not, in general, match the local minima (red) of the pinning potential (contour lines in background at the top layer). **b,** The instantaneous speed averaged over the 72 vortex lines is represented as a function of time during the positive (blue) and negative (red) half-cycles for the current amplitudes $F_m$. The peaks of the vortex lattice hopping between low velocity points are clearly visible and increased symmetrically on both sides of $t_{dep}$ = 0.25 t/period. **c,** The range of vortex excursions averaged over all vortices is plotted in the top panel as a function of current amplitude. The motion is globally proportional with the driving current but increases step by step, similar to the experiment (in insert). The numbers are labels for the mode-locking steps and perfectly match with the number of hops (b). The bottom panel shows the time-averaged pinning energy per vortex. Its dependence on the excitation amplitude reveals that, in average, the system tends to save both elastic and pinning energies when vortices are either moving inside single wells over an integer number of the lattice spacing (steps 1, 3 and 5). The steps in the curve and the local minima in $E_{pin}$ were also observed for other orientations of the driving force and different realizations of the random pinning potential (see SI 11). The initial state corresponds to one of

the available local minima of the system free-energy, prepared via a simulated annealing scheme that mimics the experimental field cooling procedure. The trajectories and energies of vortices are presented in Figure 3 with ramping up the amplitude of the driving force $F_m$.

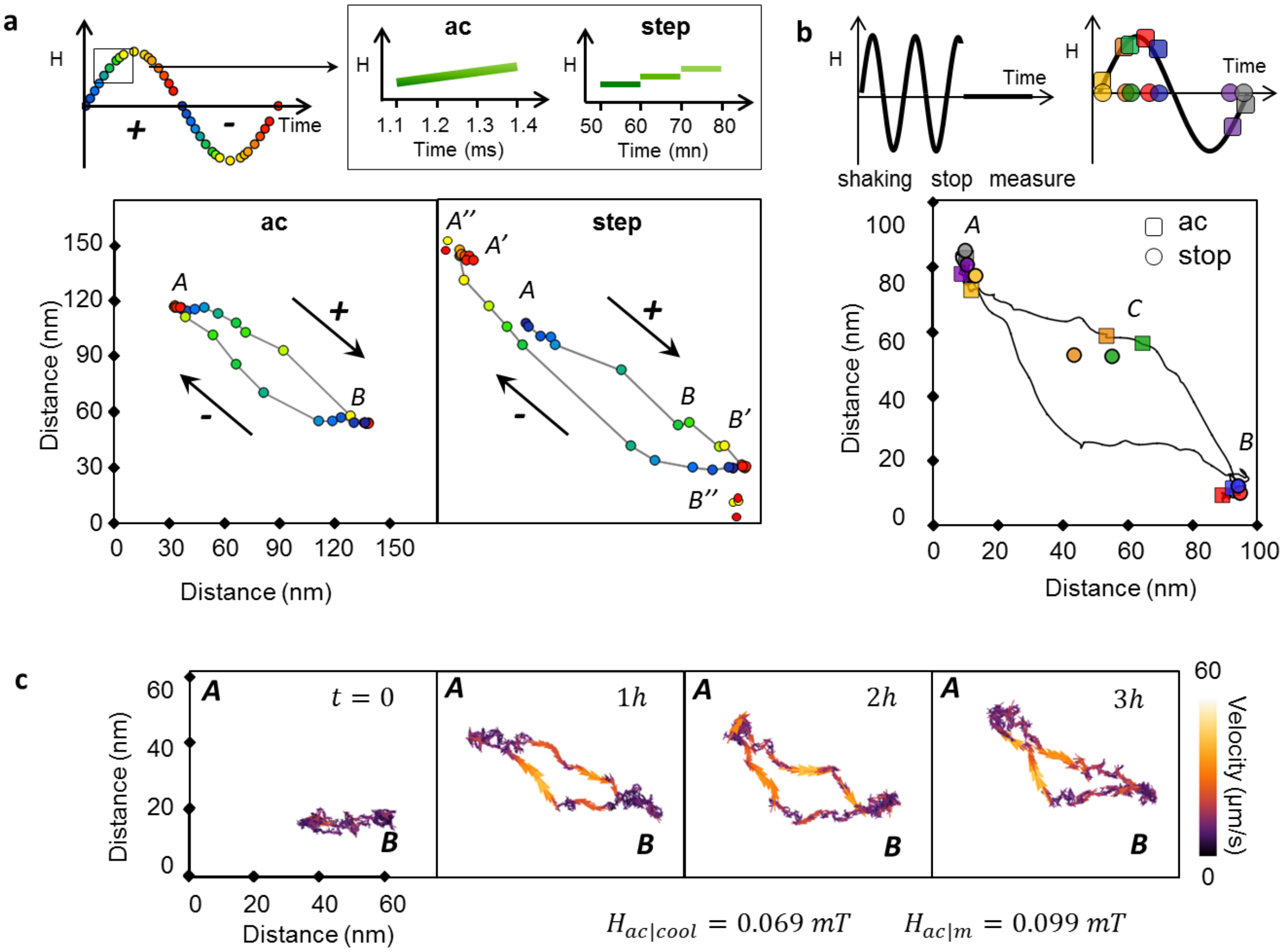

**Figure 4| Creep motion and memory effect in $NbSe_2$ at 0.5 K. a,** An ac-modulation is first applied with a period $T_{ac}$ = 13 ms (left panel), and then step by step with an equivalent period $T_{dc}$ ~ 1 h (right panel). In the slow case, an additional creep motion occurs after the high-velocity jump, leading to a larger trajectory (A'B') compared to the fast case (AB). After one hour, the vortex is observed quasi-static (B" and A" match with B' and A' respectively) (175 x 175 nm$^2$. $H_{ac|cool}/H_{ac|m}$@0.099/0.099 mT, $f_{cool}/f_m$ @77/77 Hz). **b,** The ac-excitation is successively stopped

at $t_i$ = 0.7 ms (yellow), 2.0 ms (orange), 2.2 ms (green), 3.5 ms (red), 4.0 ms (blue), 11.0 ms (violet) and 12.6 ms (grey). The vortex lattice is observed perfectly frozen, shown by the same positions adopted between ac- (square) and stop- (circle) points (see zoom). The small differences are interpreted as spatial fluctuations of the global trajectory (black line) (see SI3). (100 x 100 $nm^2$, $H_{ac|cool}/H_{ac|m}$ = 0.099/0.099 mT, $f_{cool}/f_m$ =77 Hz). **c,** After increasing the shaking, the motion of two neighbor vortices in the linear regime (1W) relaxes in the nonlinear regime (2W), reproducing the spatial shapes and low velocities of the first one (see SI9). ($H_{ac|cool}/H_{ac|m}$@0.069/0.099 mT, 62 x 75 $nm^2$).

# Stick-slip Phenomena and Memory Effects in Moving Vortex Matter

*Lise Serrier-Garcia[1*], Clécio C. de Souza Silva[2], Matias Timmermans[1], Joris Van de Vondel[1] and Victor V. Moshchalkov[1]*

[1] INPAC – Institute for Nanoscale Physics and Chemistry, KU Leuven, Celestijnenlaan 200D, B–3001 Leuven, Belgium

[2] Departamento de Fisica, Universidade Federal de Pernambuco, Cidade Universitaria, 50670-901 Recife, Pernambuco, Brazil.

* e-mail: serriergarcia.lise@fys.kuleuven.be

## A. DETALILLED EXPERIEMNTAL AND SIMULATION METHODS

### SI1. The ac-driven field-cooled vortex lattice

The vortex lattice is a perfect regular triangular Abrikosov lattice with vortex spacing $d_{v-v} = 135\ \pm 1\ nm$. The magnetic induction B is deduced from $d_{v-v} = \sqrt{2\Phi_0/\sqrt{3}B}$ where $\Phi_0$ is the flux quantum. The obtained value, B = 0.9 H, expresses a rather uniform flux distribution over the sample. Such a result indicates that order covers a larger scale as compared to the characteristic vortex-vortex interaction length $\lambda_{ab} \simeq 70\ nm$.

The structure has been well characterized and consists of Nb planes sandwiches between two Se planes [Brown 1965]. The Ginzburg-Landau coherence length $\xi_c$ perpendicular to the layers $\xi_c = 9\,nm$ is much larger than the distance between two superconducting Nb layer $l = 6\,nm$, resulting in strongly coupled layers. The ratio $\Gamma = \xi_c/\xi_{ab} = \lambda_{ab}/\lambda_c$ quantifies the anisotropy between the ab-planes and the perpendicular c-axis, leading to $\Gamma = 1$ in case of isotropic superconductors, $\Gamma_{NbSe_2} = 3.3$, $\Gamma_{YBaCuO} = 5$ and $\Gamma_{BiSrCaCuO} = 60$. The vortex lattice orientation follows the three-fold symmetry of the ab-layers. Vortex star-shapes hybridize as expected in purest $NbSe_2$ single-crystals [Suderow 2014], resulting from the in-plane gap anisotropy and charge density wave.

The vortex core motions are observed coherent and collective over tens of microns, proving large Larkin domains of a moving Bragg glass. Hence, the sinusoidal magnetic field generates vortex "edge" dynamics propagated inside the superconductor by a smooth compression of the quasi rigid vortex lattice [Campbell 1969, Brandt PRL 1991, Brandt 1992 Physica C]. In consecutive sequences of field-cooling, the vortex lattice occupies different positions, providing a multitude of energetically quasi-equivalent metastable states according to a weak pinning situation [Ovchinnikov 1983].

In summary of Fig. 1 and 2 in the main text, vortices are driven by an ac-magnetic field resulting in a collective linear response in space and time. However, trajectories reveal that the motion is not coherent with the ac-magnetic field while prepared or shaken with an ac-excitation above a threshold value. Vortices in $NbSe_2$ were already observed to travel against the driving force [Raes 2014, Timmermans 2014]. Particularly, the motion along one of the vortex lattice axis ($\theta = 0$) was interpreted in terms of glass motion [Le Doussal 1998, Olson PRB 2000] in Ref. [Troyanoski 1999] and ab-plane gap-anisotropy in Ref. [Timmermans 2014]. Nevertheless, trajectories in the nonlinear regime show different preferential directions ($\theta = +0.28\,\pi$), rejecting the static or dynamic potentials suggested in these two articles.

**SI2. The synchronized Lazy-Fisherman spectroscopy to obtain real-time trajectory of vortices [Timmermans 2014]**

**a. The synchronized Lazy-Fisherman spectroscopy.** In the "Lazy-fisherman" method [Kohen 2005] the STM tip is kept at a fixed position while vortices pass underneath. The observation of a moving vortex is given by a sudden change of the tunneling current. Since, the electronic density of state differs between the normal core and the superconducting state at a bias voltage U lower than the equivalent superconducting gap Δ (Figure S2a). In order to control the distance between the tip and the surface, the scan voltage is selected outside the gap range (U>Δ) where the current is equal in both normal and superconducting states. During spectroscopy, the feedback loop is turned off and then the current is measured at a bias voltage inside the gap (U<Δ). As such, this method detects motion of a vortex at one specific location. Achieving a complete overview of spatial motion, the STM tip (fisherman) is positioned throughout the scan area, pixel by pixel. At every point the fisherman waits while measuring the tunnel current during at least one period of the ac oscillation (see markers in Figure S2a). Synchronizing the lazy-fisherman spectra with the ac-excitation of the magnetic field, the obtained current variations are linked with the ac-period. Finally, for each step in time, we extract the current value and place it in a spatial map at the corresponding pixels coordinates (Figure S2b).

**b. Vortex trajectory.** On the contrary to the technique measuring the tunneling current with the feedback loop on in ref [Timmermans 2014], the distance between the tip and the sample is kept constant in the spectroscopy mode, and the vortex cores can be visualized in the I-maps (Fig. S2c-e). From the set of 520 current maps, we extract the vortex positions $(x(t_i),y(t_i))$ by fitting the vortex shape with a 2D Gaussian function. A trajectory is obtained by tracing all these points in one spatial map. Instantaneous velocities are numerically derivated from two consecutive points in time.

This technique is based on reproducibility of the motion over the complete measurement, resulting in identical vortex shapes in static and dynamic. However, during the jump, vortex shapes appear elongated, which can be due to two possible reasons (Fig. S2b). First, vortices move faster than the speed resolution (~ 40µm/s) fixed by the spatial and time parameters. Secondly, vortices can jump at different depinning time, resulting in similar smearing.

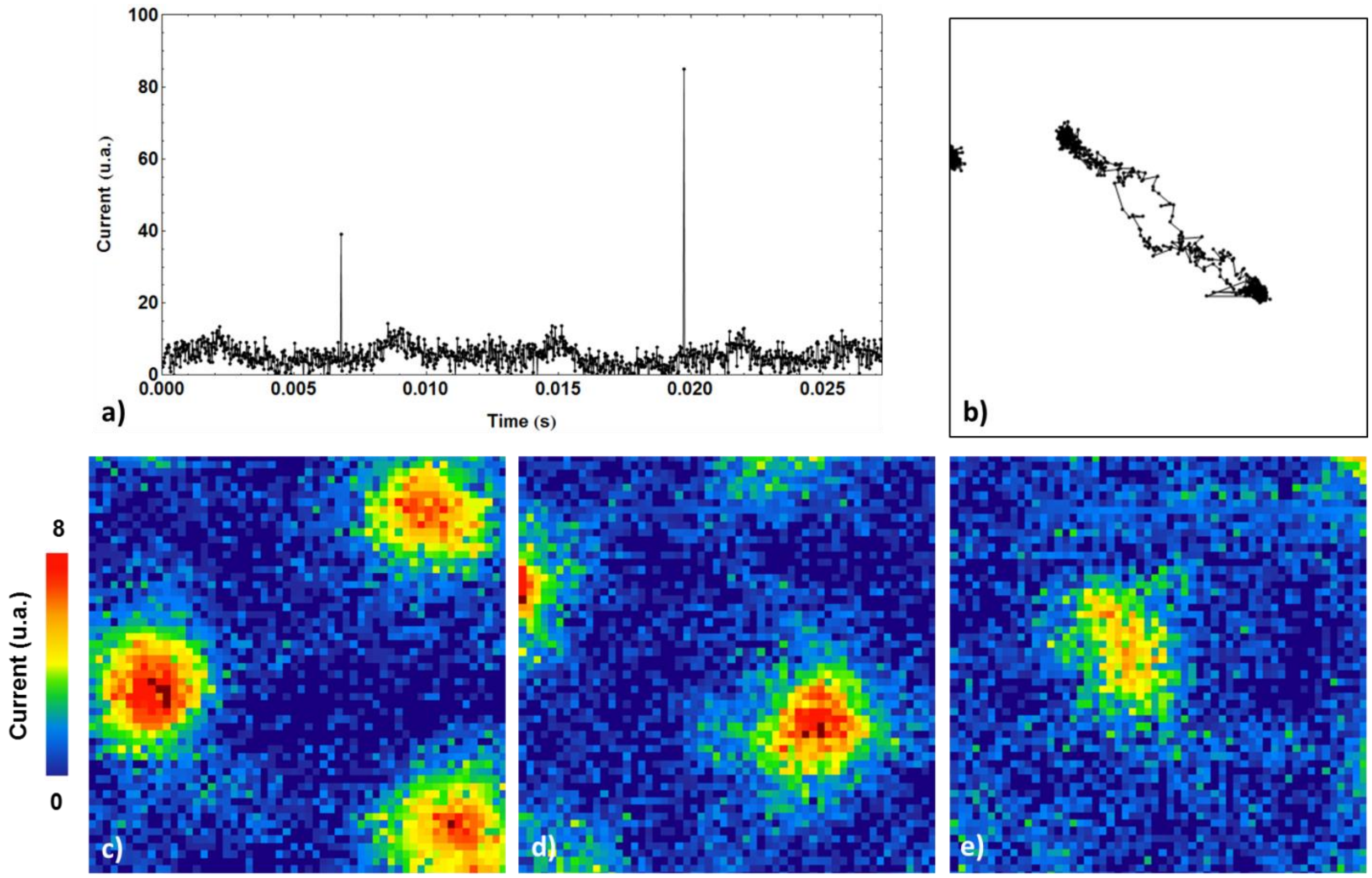

**Figure S2| The synchronized Lazy-Fisherman spectroscopy. a,** An example of current spectroscopy is shown during two ac-periodes measured with a bias voltage of 1 mV. The current values vary between 3 a.u. and 8 a.u. corresponding to the superconducting state between vortices and the normal state at vortex cores respectively. The ac-cycles are noted by an electronic pulse in the current circuit leading to two sharp spikes above 40 a.u.. The modulations in the curves correspond to vortices passing through the pixel. In this example, two broad peaks appear per period, one for forward (positive half ac-cycle) and the second for backward (negative half ac-cycle). **c, d, e,** All vortex positions are extracted from the current maps to construct complete trajectories (b). Vortex shapes (red spot) in dynamic look similar to the static situation (c) for the creep motion (d) while smeared for the flux-flow (e). (175 x 175 nm²).

**SI3. Further detail on the theoretical modelling**

We approach theoretically the dynamics of a 3D vortex system oriented along the c-axis of an anisotropic superconductor by using Langevin-Bardeen-Stephen equations of motion [van Otello 1998]:

$$\eta \frac{\partial \vec{r_i}(z,t)}{\partial t} = -\frac{\delta \mathcal{F}}{\delta \vec{r_i}} + \phi_0 \vec{j}(z,t) \times \hat{z} + \vec{\Gamma_i}(z,t), \qquad (Eq\ 1)$$

where $\eta$ is the Bardeen-Stephen drag coefficient, $\mathcal{F}$ is the free-energy density describing the elastic properties of the vortex system and its interaction with pinning, $\vec{j}$ is the applied current density, and $\vec{\Gamma_i}$ is the thermal random force.

The free-energy density is modeled as:

$$\mathcal{F} = \sum_i [U_p(\vec{r_i}, t) + \frac{\epsilon_v}{2}\left|\frac{\partial \vec{r_i}}{\partial z}\right|^2 + \sum_{j>i} U_{int}(\vec{r_i} - \vec{r_j})], \qquad (Eq\ 2)$$

where $\vec{r_i}(z)$ describes the 2D position of the $i$-th vortex. For simplicity, the effective vortex line tension is estimated within the local approximation as $\epsilon_v = \varepsilon^2 \epsilon_0 (1-b)$, where $\varepsilon = \lambda_{ab}/\lambda_c$ is the anisotropy parameter, $\epsilon_0 = \phi_0^2/4\pi\mu_0\lambda_{ab}^2$, and $b = B/B_{c2} = 0.032$ is the reduced flux density [Kierfeld 2004]. The pair interaction energy per unit length is modeled within the London limit as a modified Bessel potential, $U_{int}(r_{ij}) = \epsilon_0 K_0(r_{ij}/\lambda_{ab})$. In order to boost the simulation speed, we calculate the total interaction potential acting on a given segment of the $i$-th vortex as if the corresponding segments of all other vortices were fixed at their perfect lattice position. This so-called cage model provides a good approximation for the cases where the weakness of the quenched and thermal disorders disturbs the vortex lattice only slightly [Frey 1994, Ertas 1996]. Here, since the system is constantly driven out of equilibrium, one needs to calculate self-consistently the position of the vortex lattice center-of-mass [de Souza Silva 2002] in order to accurately estimate the attraction center of the cage potential for each vortex.

For each z plane, the pinning potential $U_p(\vec{r_i}, z)$ is evaluated on a two-dimensional grid of square cells of size $0.2\,\xi_{ab}$ as the convolution of a random pinning function, sampled from a Gaussian distribution, and the form factor of the vortex core, which, within the Ginzburg-Landau theory, is well approximated by $f(r) = \xi_v^2/(r^2 + \xi_v^2)$, with $\xi_v \simeq \sqrt{2}\xi_{ab}$ [Blatter 1994]. The z axis is coarse-grained over a scale $1.875\,\xi_{ab}$, in a way that each vortex segment represents effectively several ab planes. In order to solve the problem numerically, we further discretize Eq 1 in time. The material parameters were chosen close to the typical values for $NbSe_2$ evaluated at T = 0.5 K: $\xi_{ab} = 10\,nm$, $\lambda_{ab} = 150\,nm$, $\varepsilon = 0.3$, $\epsilon_0 = 12\,pN$ and $\eta = 1.36 \times 10^{-7}\,Ns/m^2$. All simulations were performed for a vortex density corresponding to B = 100 mT, similar to the dc magnetic induction used in the experiments. For the chosen simulation cell size $L_x \times L_y \times L_z = 120\,\xi_{ab} \times 117\,\xi_{ab} \times 240\,\xi_{ab}$, this corresponds to 72 vortex lines, each one comprising 128 vortex segments (9216 vortex segments in total).

## B. ADDITIONAL EXPERIMENT RESULTS

### SI4. Investigations of spatial fluctuations during motion

We investigated relaxation in the $10^3$s time range. In Fig. S4, trajectories were measured several times after preparing the system. The identical features show a deterministic motion over several hours corresponding to four million ac-cycles. However, different signatures in trajectories spaced by two hours reveal spatial fluctuations with time and reflect intrinsic behaviors of creep motion with thermal or quantum fluctuations.

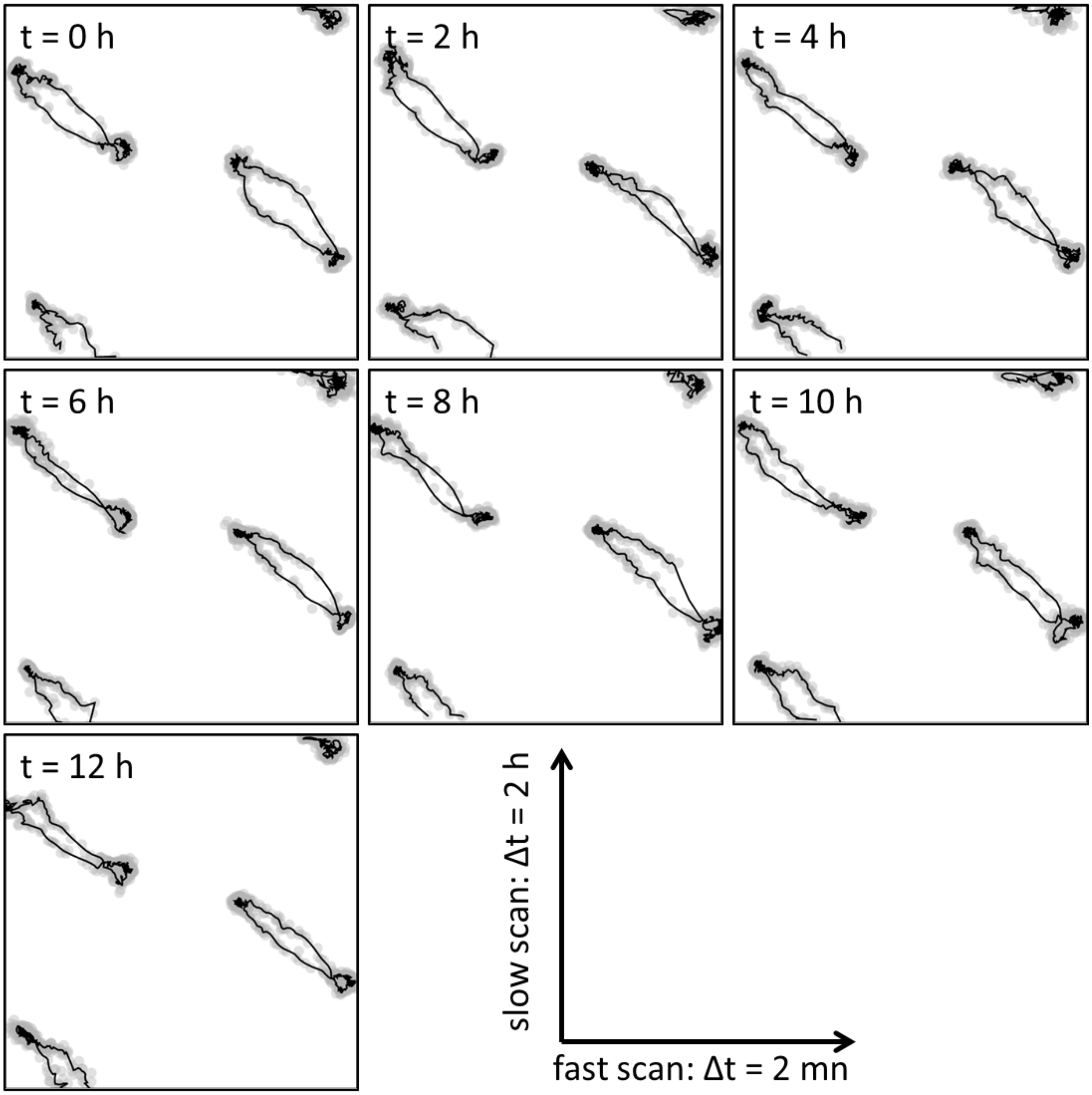

**Figure S4| Spatial fluctuations vs time.** Reproducibility of motion is investigated by repeating seven times the measurement. Trajectories are globally identical during 14 h although some spatial fluctuations are bigger than the error bar of the position (in grey). (200 x 200 nm², $H_{ac|cool}/H_{ac|m}$@0.082/0.099 mT).

**SI5. Frequency study**

According to the common electric field criterion (Ec ~ 10-100 μV/m [Matsushita 2014]), the high velocity of the vortex jump corresponds to a dissipative value of flux-flow (vB ~ 8-30 μV/m) while low velocities match with creep value (vB < 1 μV/m).

In the range of 15 Hz to 121 Hz, the high velocity peak occurs at the frequency-independent time $t_{dep} = 0.18\ T_{ac}$, which confirms that depinning is governed by the instantaneous amplitude of the external excitation. It is well-established that ohmic losses due to a viscous medium depend on the ac-frequency [Schmidt 1972, Sonin 1990, Geshkenbein Physica C 1991]. However, in our case they seem negligible. One possible explanation is that the flow-velocity range concerns only the jump and covers less than 5% of the ac-cycle, while the rest is frequency independent as expected at low velocity [Brandt PRL 1991].

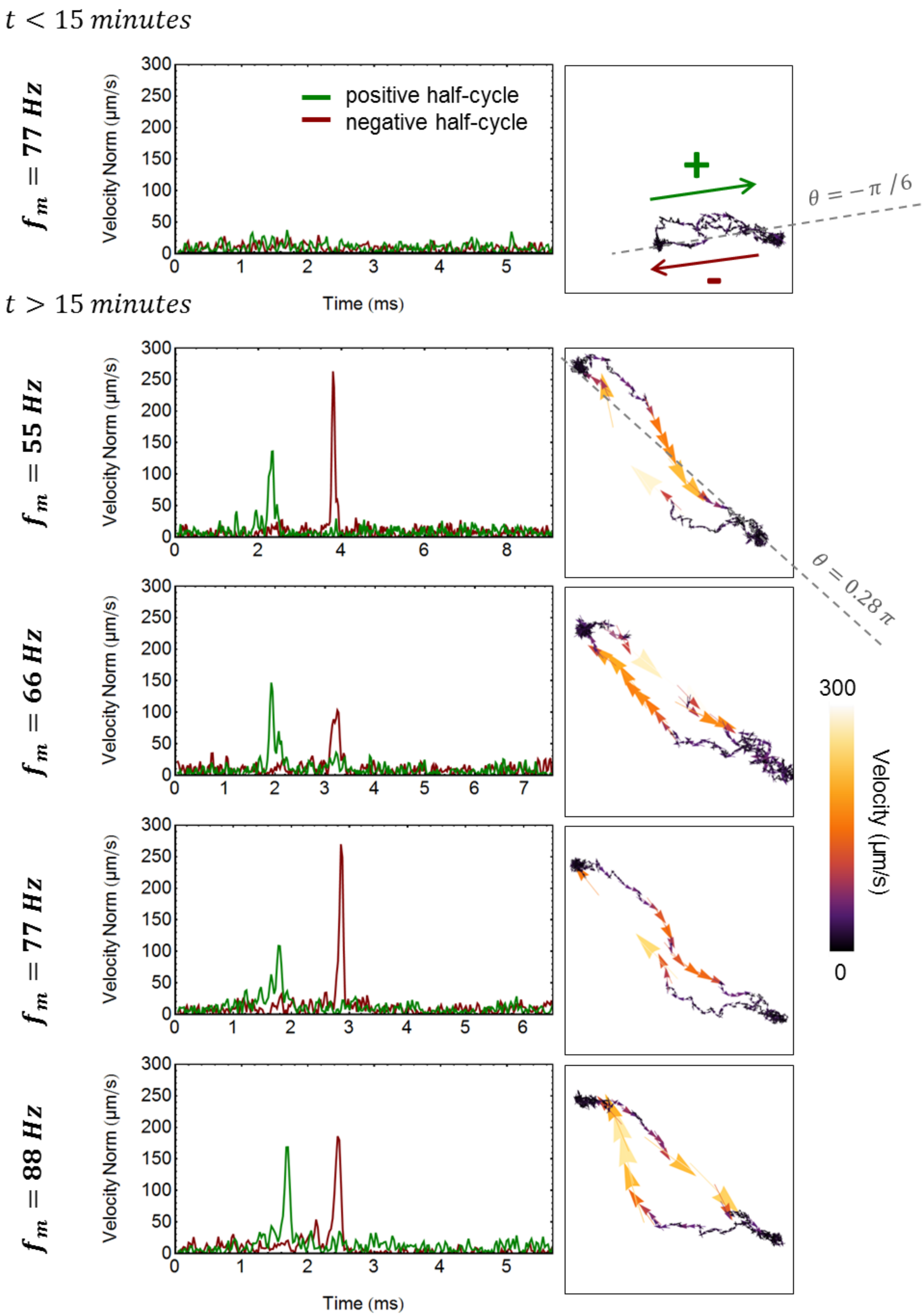

$t < 15\ minutes$
$f_m = 77\ Hz$
$t > 15\ minutes$
$f_m = 55\ Hz$
$f_m = 66\ Hz$
$f_m = 77\ Hz$
$f_m = 88\ Hz$
Velocity Norm (μm/s)
Time (ms)
positive half-cycle
negative half-cycle
+
-
$\theta = -\pi/6$
$\theta = 0.28\,\pi$
Velocity (μm/s)
300
0

a.

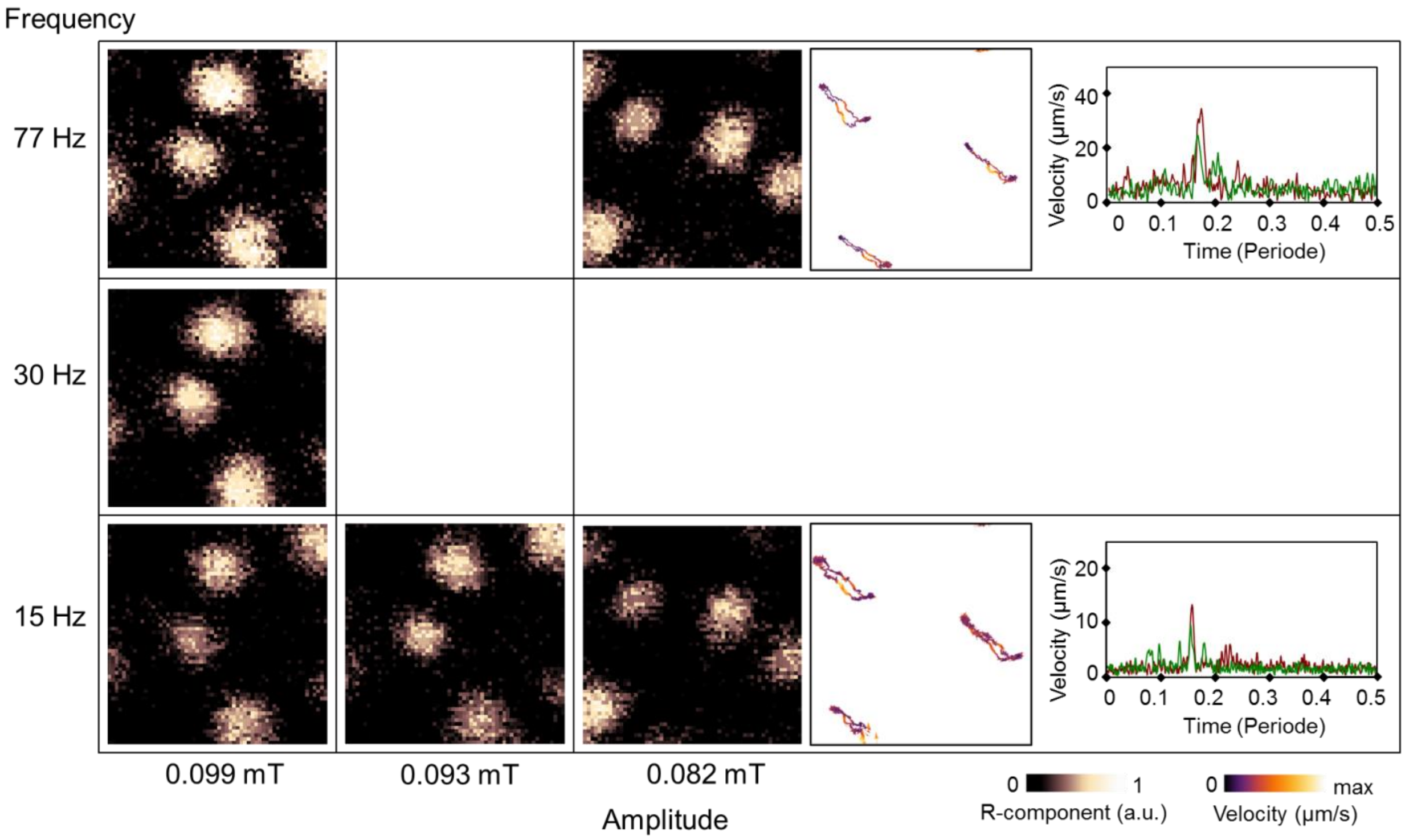

b.

**Figure S5| Trajectory vs frequency. a,** During the first fifteen minutes after increasing the shaking from 0.093 mT to 0.099 mT, the vortex lattice follows a one-potential well (1W) trajectory, i.e. the linear regime, before suddenly relaxing in a two-potential well (2W) path. Then, this nonlinear identical trajectory does not evolve and, in particular, it is identical for different frequencies $f_{|m}$ = 55, 66, 77 and 88 Hz. In the velocity graphs, the high-velocity peaks appear in all cases at $t_{dep+} = 0.14$ $T_{ac}$ (green) and $t_{dep-}$ =0.22 $T_{ac}$ (red) ($H_{ac|cool}/H_{ac|m}$@0.093/0.099 mT, $f_{cool}$@77Hz). **b,** Similar results are obtained in another experiment at $f_{|m}$ = 15, 30, and 77 Hz. As shown in the R-component maps (see technique in

[Timmermans 2014]), the oscillation amplitude is frequency-independent and increases with the ac magnetic field amplitude. ($H_{ac|cool}$@0.099mT, $f_{cool}$@77Hz)

**SI6. Hysteresis of the depinning transition**

In figure S6, amplitude of a triangular ac-excitation was varied up and down between 0.082 mT and 0.120 mT after cooling the system without $H_{ac}$ (i.e. $H_{ac|cool}$ = 0). In order to get fast overviews of motion, we used another technique described in detail in [Timmermans 2014]. In this technique, the tunneling current is combined in real-time with an electronic sinusoidal signal of the same frequency. The resulting R-component, represented in the top panel of Figure S6, gives information on amplitude and direction variations. The R-component (a) and trajectories (c) measured with increasing shaking show one low-velocity point (1W) paths up to 0.099 mT, followed by the transition $H_{1W\rightarrow 2W}$ to the two low-velocity point (2W) regime between 0.099 mT and 0.104 mT. I point out that the $n$W correspond to the different $n$-steps obtained by the simulations. At 0.120 mT, the two maps show relaxation of the 2W motion with time. During the decrease of $H_{ac|m}$, the transition $H_{2W\rightarrow 1W}$ to the 1W regime occurs between 0.093 mT and 0.088 mT. This value is in good agreement with the depinning amplitude $H_{dep}$ (0.080 < $H_{dep}$ < 0.086 mT) noted by the peak in the 2W velocity graphs (see the main text). The small difference between $H_{1W\rightarrow 2W}$ and $H_{dep}$ is probably a consequence of the dynamical vortex-lattice configuration or the appearance of the bending force $f_{bend} = H_{1W\rightarrow 2W} - H_{dep}$. Moreover, point A is more attractive than point B, revealed by an asymmetry of $H_{dep}(t_+)$ and $H_{dep}(t_-)$ in the velocity graphs (see SI7) and, as a result, an additional hysteresis of the transitions $H_{1W\rightarrow 2W}$ and $H_{2W\rightarrow 1W}$ with sweeping $H_{ac|m}$.

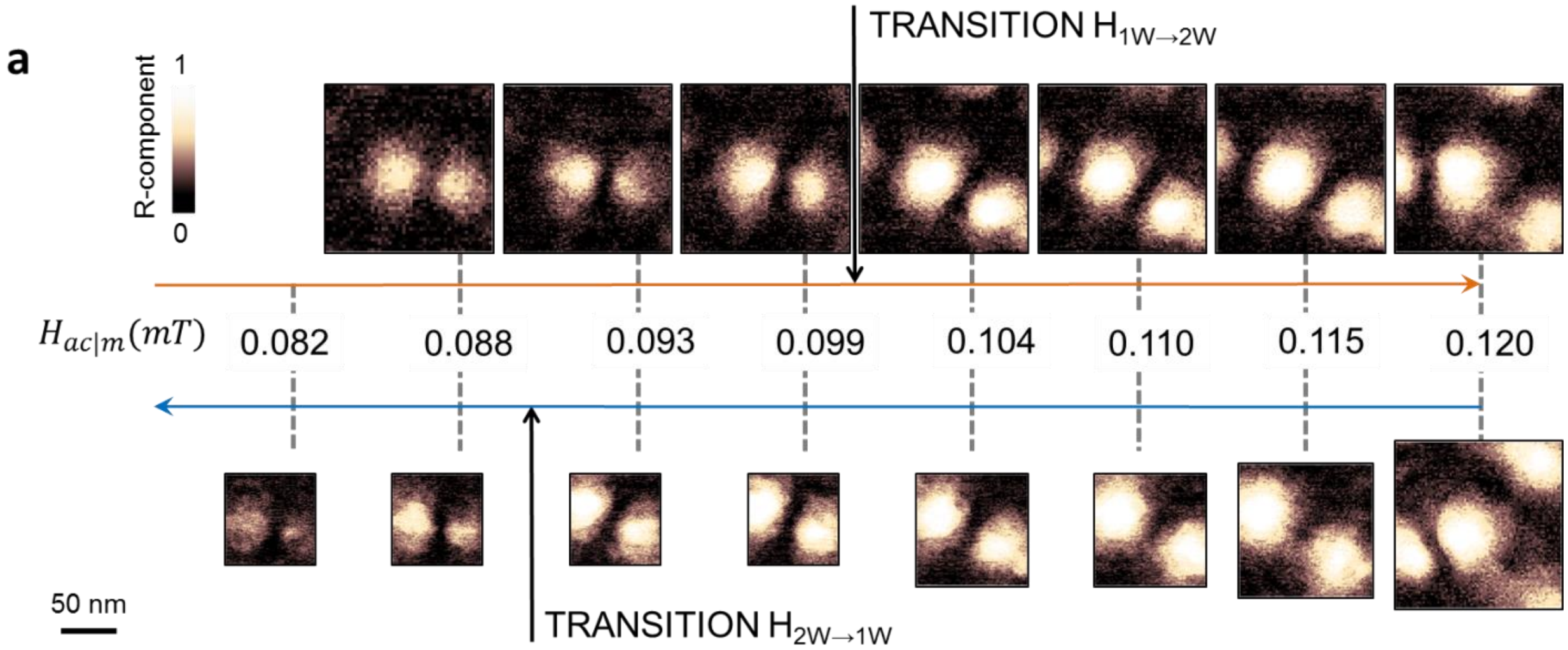

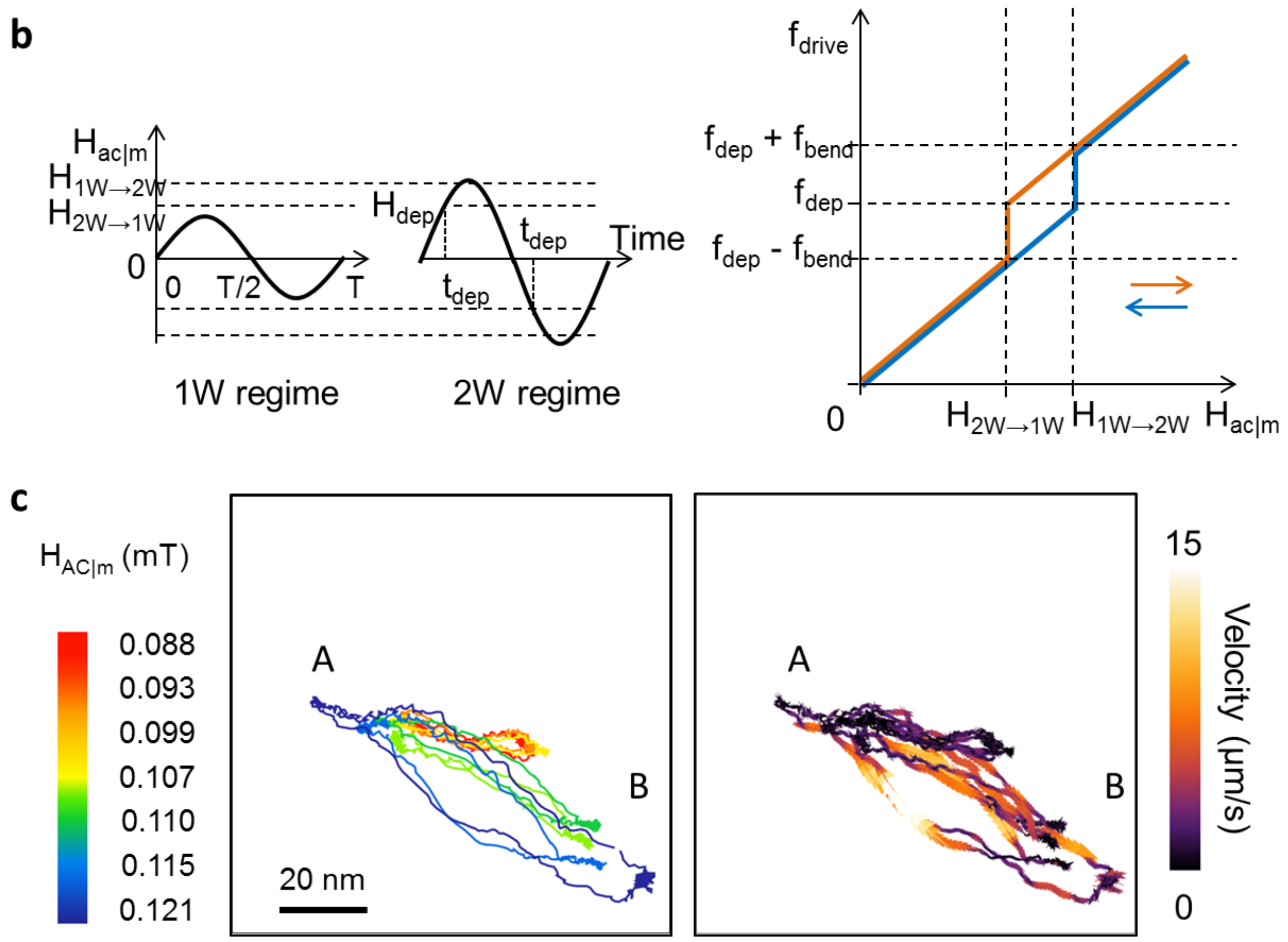

**Figure S6| Transitions between one-potential well (1W) and two-potential well (2W) trajectories vs $H_{ac|m}$. a,** The R-component maps show a hysteretic transition with the amplitude of ac-excitation. **b,** In the 1W regime, the driving force works against the pinning force. Above $H_{1W\rightarrow2W}$, the additional force $f_{bend}$ reduces $H_{2W\rightarrow1W}$. In the 2W regime, the jump value $H_{dep}$ of trajectories equals the transition value $H_{2W\rightarrow1W}$. **c,** Trajectories show the transition from the 1W to the 2W regimes with increasing $H_{ac|m}$ instead of $H_{ac|cool}$ (Figure 2a of the main text). ($H_{ac|cool}$@0 mT, triangular ac-excitation).

**SI7. Low-velocity points in the stick-slip motion**

In figure S7, thirty minutes after cooling the system, the vortex lattice jumped from trajectory A'B' to trajectory AB following the driving vector $(D, \theta) = (a_0/6, +0.28\,\pi)$. The first trajectory is presented in dashed lines as a guide for eyes. In the second trajectory presented here, a third low velocity point C appear at the location previously occupied by A'. However this memory effect is not visible in the velocity graph where the two jumps revealed by two high-velocity peaks appears symmetric at $t_{\alpha,\beta} = t_{jump} \mp t_{sym}$ with $t_{sym} = 0.6\,ms$ (this corresponding to $H_{\alpha,\beta} = H_{jump} \mp H_{sym}$ where $H_{sym} = 0.012\,mT$. $H_{sym} = 0.018\,mT$ was obtained in the case of a triangular signal in Figure S9). The symmetric shift around $t_{jump}$ reveals that the depinning of the vortex bundle from the linear regime with increasing $H_{ac|m}$ is independent of the jump $H_{jump}$, $H_\alpha$ and $H_\beta$ of step 2 and step 3 of the nonlinear regime (see Fig. 3 in the main text).

Additionally, an asymmetry exists between the positive $(t_+)$ and negative $(t_-)$ half ac-cycles, with $t_\pm = t_{dep} \mp t_{asym}$, reflecting an unequal balance between forward and backward (equal to 0.005 mT in Figure SI7).

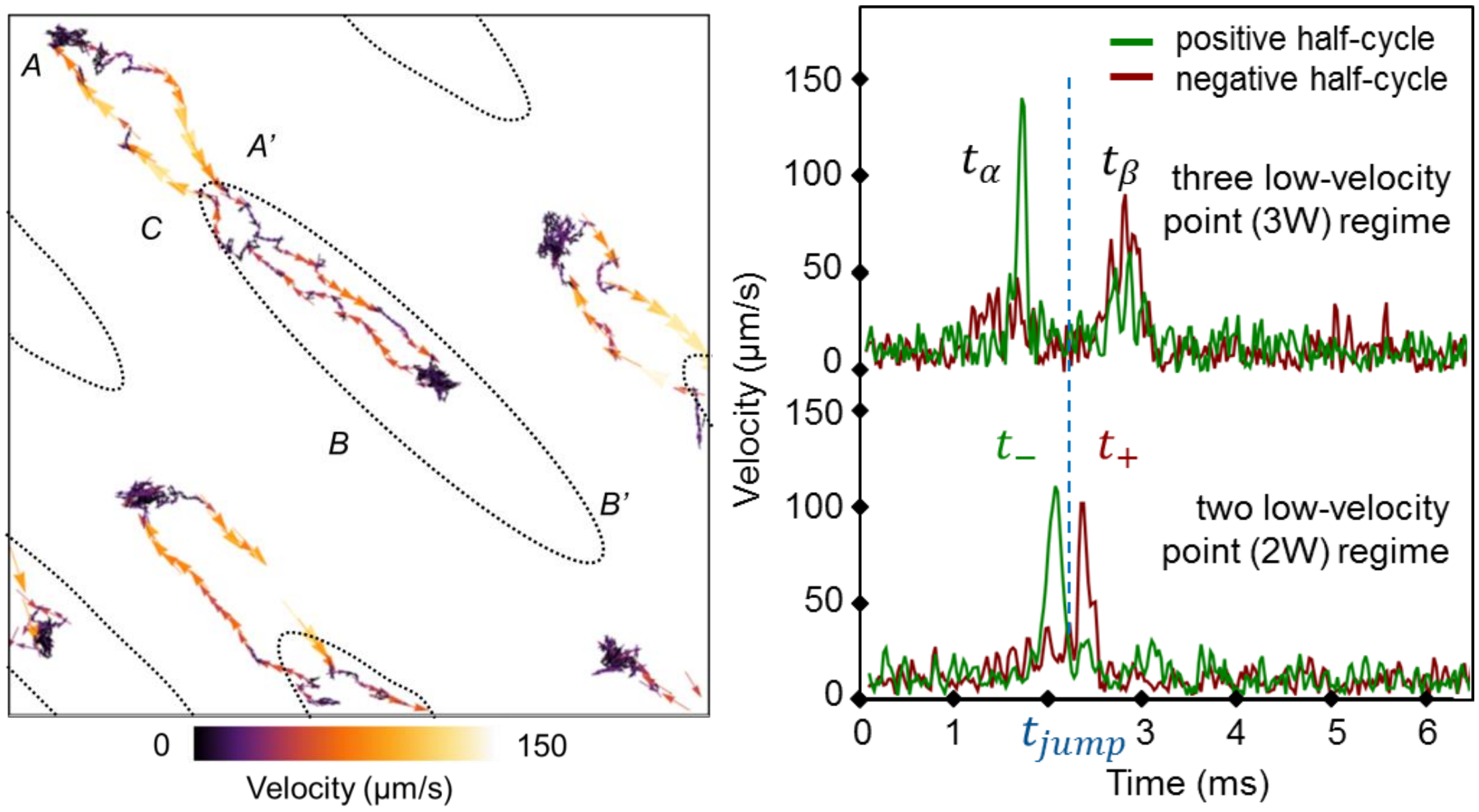

**Figure S7| Formation of a low-velocity point in the two well (2W) regime.** A vortex lattice relaxed few minutes after cooling with shaking, leading to a three low-velocity point (3W) regime. After a sudden drift of the vortex lattice, the low-velocity point A' was kept as a low-velocity point C. (200 x 200 nm² $H_{ac|cool}/H_{ac|m}$@0.099/0.099 mT). The corresponding velocity graph is presented in comparison with the velocity graph of a trajectory in a 2W regime.

## SI8. Additional results on the system relaxation

Figure S8 complete Figure 4c (see main text) presenting another experiment with trajectories measured at different amplitude of shaking (a). In the R-component map (b) measured before getting trajectory n°4 at 0.121 mT (a), a black spot seems cut at the scan line noted by the arrow. This reflects that the motion suddenly changed at the corresponding time in scan (following the low scan direction). After increasing amplitude of shaking form 0.077 mT to 0.121 mT, the top part A of trajectory was reached in few seconds (see the white spot A' revealed by the first lines of scan) while the bottom part B was accessed only one to two minutes later.

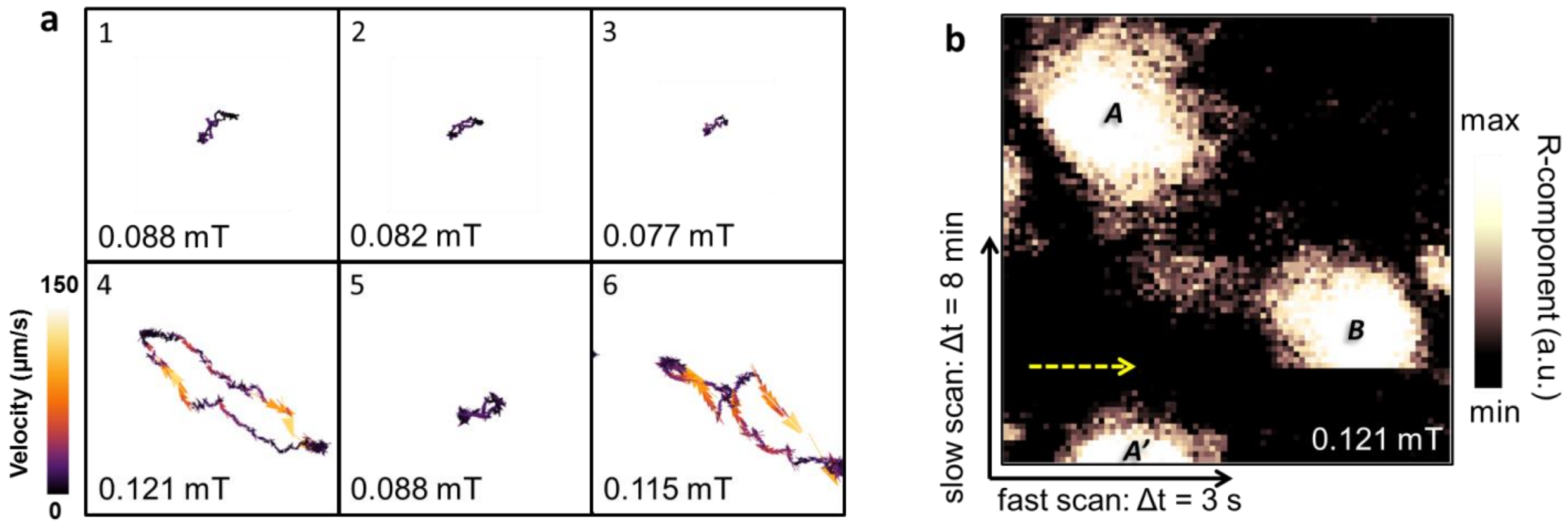

**Figure S8| Relaxation versus regimes switching. a,** Reversibility of the one low-velocity point regime to the two low velocity point one versus amplitude of shaking. **b,** The transition from the 1P to the 3P regimes of Figure 4b (see main text) is complete after around 10 000 ac-cycles. (150 x 150 nm$^2$).

**SI9. Additional information Figure 4a**

Two neighboring vortices are following the linear regime (1W) before relaxing in the nonlinear (2W) regime (see main text).

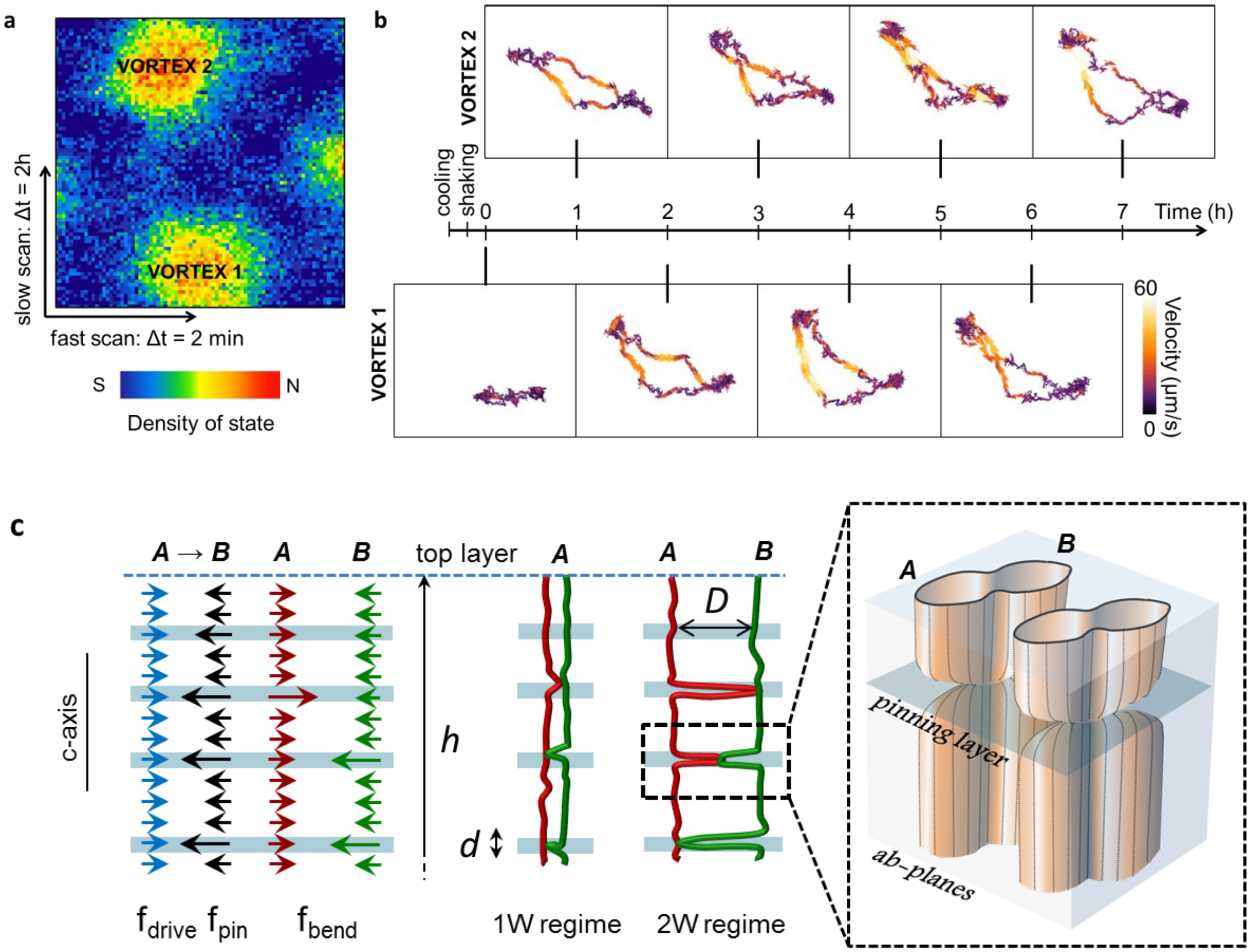

**Figure S9| Relaxation versus time after increasing amplitude of shaking . a, b,** The two neighboring vortices of Figure 4a (see main text) are probed during 6 hours after increasing the excitation from 0.069 to 0.099 mT. (**a,** dIdV@1mV, 200 x 200 nm². **b,** 62 x 75 nm²). **c**, Vortex profiles along the c-direction illustrate the vortex configurations in the two different regimes (1W and 2W), at t = 0 (green) and t = T/2 (red). The different forces applied to the vortex lattice in the potential wells A and B are represented for the nonlinear 2W regime. The free straight vortex segments move coherently in the ab-planes while bending acts in the c-direction. The zoom shows the 3D trajectory of two neighbor vortices bending collectively around one pinning segment.

The linear path (t = 0) appears in the nonlinear motion after relaxation of the system (see also S5 and SI8-9). As a local attractor along the linear motion direction was not observed in other part of the nonlinear trajectories, this reflects a history-dependence of the motions which cannot be explained using conventional pinning arguments and a more elaborate discussion is needed. We believe that a rigid vortex lattice moves coherently in the ab-planes and is attracted by certain strongly pinned vortex segments at other coordinates z. In layered superconductors, it is worth considering the flexibility of the vortex lattice compared to the usual two-dimensional assembly of rigid tubes in conventional superconductors. Hence, the length $d$ of the tilt (see Fig. S8.1) is reduced due to the layer nature of the material, determined by the anisotropy strength $\Gamma$, $\tan(\gamma) \leq \Gamma$ where $\gamma$ is the bending angle [Brandt PRL 1992]. In case of $NbSe_2$, $\Gamma$ is equal to 3.3. According to the prediction, $d$ is estimated ~ 100 nm for a displacement of AB $\cong$ $a_0$ (maximum amplitude of jump in trajectories); this small value allows each vortex to bend $10^4$ times over the sample thickness $h$.

## B. ADDITIONAL SIMULATION RESULTS

### SI10. Average contribution of the pinning, interaction and line energies to the total free energy

In Fig. S10, we show the time-averaged pinning, line, and interaction energies per vortex as functions of the drive amplitude: $\langle E_{pin} \rangle$, $\langle E_{piline} \rangle$, and $\langle E_{int} \rangle$, corresponding respectively to the first, second and third terms in the right-hand side of Eq. 2 in SI3 integrated along the z axis and averaged over time and the 72 vortices. The total averaged free energy is also shown (top panel). As clearly illustrated in Fig. S10, the main contribution to the total free energy of the system comes from the mean pinning energy. Indeed, precisely because the pinning potential is so weak, deformations of the vortex lattice, within the ab planes and along the c axis, are very small, leading to the observed small, almost negligible, contributions of these elastic energies. These results justify self-consistently all approximations used in our model, which rely on small deformations of the vortex lattice.

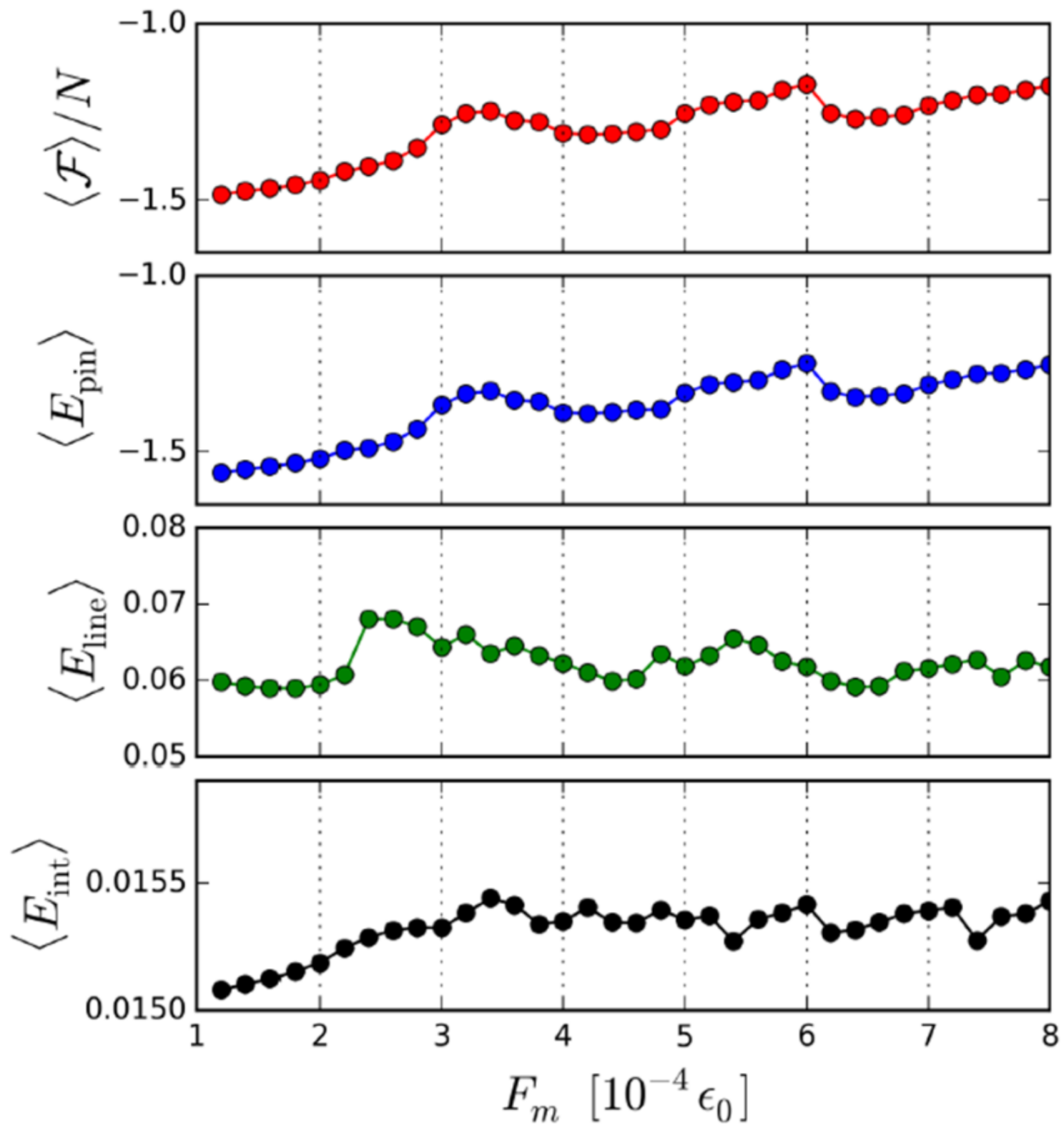

**Figure S10| Time-averaged pinning (blue), line (green) and interaction (black) energies per vortex as functions of the excitation amplitude.** The total time-averaged free energy per

vortex (red) is shown in the top panel. Energies are in units of $\epsilon_0 \xi_{ab} = 1.9 \times 10^{-19} J$. All parameters are the same used for the data presented in Fig. 3 of the main text.

### SI11. Vortex excursions for different drive orientations

In Fig. S11, we present simulation results for different orientations (θ) of the driving force. All other parameters are the same used in the main text. These results provide further evidence that vortex excursions near a lattice vector of the ideal vortex lattice, $\overrightarrow{R_{m,n}} = m\,\overrightarrow{a_1} + n\,\overrightarrow{a_2}$, where $\overrightarrow{a_1} = (1,0)a_0$ and $\overrightarrow{a_2} = (1, \sqrt{3})/2\,a_0$, are dynamically favored, resulting in the observed lock-in steps. For instance, for θ = 60°, motions at the vectors $\overrightarrow{R_{exc}} \simeq \overrightarrow{R_{1,0}}$ and $\overrightarrow{R_{exc}} \simeq \overrightarrow{R_{2,0}}$ are clearly favored, while for θ = 30° a step appears around $\overrightarrow{R_{exc}} \simeq \overrightarrow{R_{1,1}}$. For θ = 45° the mean excursion distorts slightly from the drive direction in order to favor a nearby lattice vector, inducing a plateau around $\overrightarrow{R_{exc}} \simeq \overrightarrow{R_{2,1}}$. In the insets, the trajectories for a particular vortex in different situations are shown. Notice that all observed steps corresponding to lattice vectors result in local minima in the time-average mean pinning energy as a function of drive amplitude, $\langle E_{pin}\rangle(F_m)$. Intermediate steps, at excursions not coinciding with lattice vectors, are also observed for all drive directions. However, these lock-in phases do not lead to local minima in $\langle E_{pin}\rangle(F_m)$.

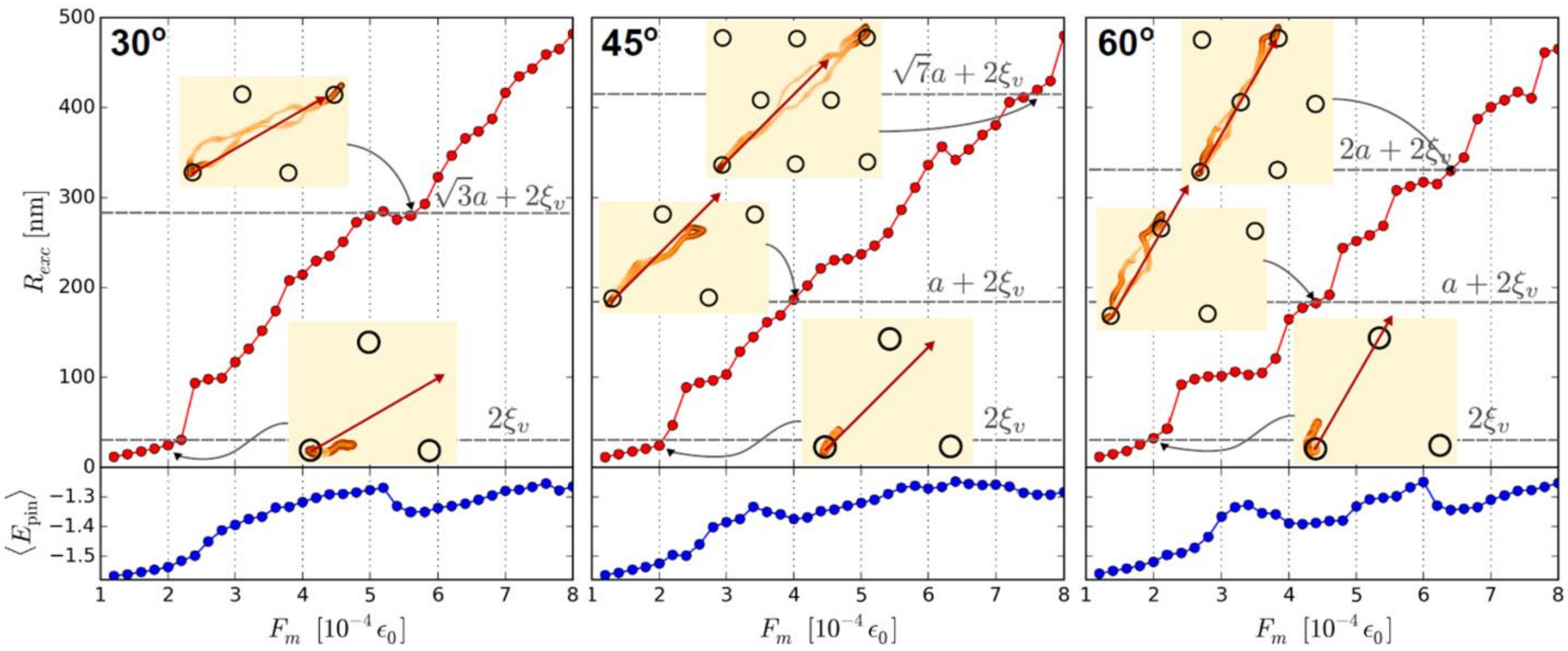

**Figure S11| Mean excursion range as a function of excitation amplitude for three different orientations of the drive (indicated in the upper left corner of each panel and by red arrows in the insets) with respect to the x axis.** All parameters are the same used for the data presented in the main text. The data for 60°, also shown in the main text, is reproduced here for comparison. The trajectories of a particular vortex (graded orange lines) and the positions of nearby vortices at the beginning of the drive cycle (open circles) in different situations (indicated by the curved arrows) are shown in the insets. In the trajectories, brighter shades indicate higher vortex speeds.

## SI12. Vortex excursions for different realizations of the random pinning potential

We repeated the simulations for several different realizations of the random pinning potential with the ac drive oriented at θ = 60°. Most of the runs exhibit similar features as those shown in Fig. 3 of the main text. A representative set of the results is depicted in Fig. S12. For instance, the data shown in the first (leftmost) panel present a similar step structure as that shown in Fig. 3 (see also Fig. S11), but with steps occurring at different drive amplitude values. For other runs, some steps are missing and sometimes additional steps appear. However, in all cases, we have found steps at the main lattice vectors $\overrightarrow{R_{exc}} \simeq \overrightarrow{R_{1,0}}$ and $\overrightarrow{R_{exc}} \simeq \overrightarrow{R_{2,0}}$, although somewhat shrunk at some instances. We have also observed in a few cases a sudden drop in $\overrightarrow{R_{exc}}(F_m)$. These were caused by an accommodation of the vortex lattice in a lower local minimum of the pinning potential. The sensitivity of the above results with respect to the different realizations of the random pinning potential is related to the small size (72 vortices) of the simulated vortex system, which renders it more sensible to fluctuations in the pinning distribution.

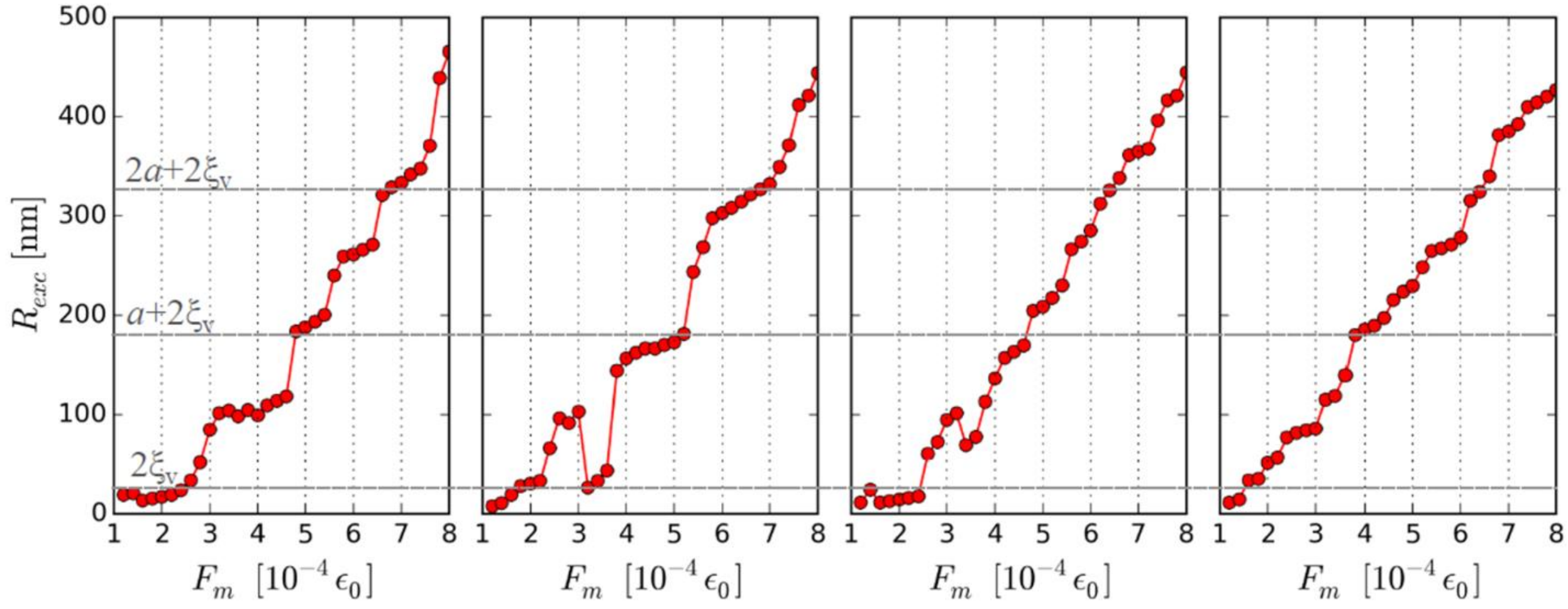

**Figure S12| Mean excursion range as a function of excitation amplitude for four different realizations of the random pinning potential.** All parameters are the same used in Fig. 3 of the main text.

**References**

Brown, B. E., Beerntsein, D. E., Layer structure polytypism among niobium and tantalum selenides, *Acta Cryst.* **18** 31-36 (1965)

Suderow, H., Guillamón, I., Rodrigo, J. G., Vieira, S., Imaging superconducting vortex core and lattice with the scanning tunneling microscope, *Supercond. Sci. Technol.* **27** 063001 (2014)

Brandt, E. H., Penetration of magnetic ac fields into type-II superconductors, *Phy. Rev. Lett.* **67** 2219-2222 (1991)

Campbell, A. M., The response of pinned flux vortices to low-frequency fields, *J. Phys. C* **2** 1492-1501 (1969)

Brandt, E. H., Flux line lattice in high-$T_c$, superconductors: anisotropy, elasticity, fluctuation, thermal depinning, AC penetration and susceptibility, *Physica C* **195** 1-27 (1992)

Ovchinnikov, Yu. N., Formation of metastable states at small-size defects and pinning in type-ll superconductors, *Sov. Phys. JETP* **57** 136-141 (1983)

Olson, C. J., Reichardt, C., Transverse depinning in strongly driven vortex lattices with disorder, *Phys. Rev. B* **61** R3811-R3814 (2000)

Kohen, A., Cren, T., Proslier, Th., Noat, Y., Sacks, W., Roditchev, D., Giubileo, F., Bobba, F., Cucolo, A. M., Zhigadlo, S.M., et al. Superconducting vortex profile from fixed point measurements the "Lazy Fisherman" tunneling microscopy method, *Appl. Phys. Lett.* **86** 212503 (2005).

Sonin, E. B., Tagantsev, A. K., Traito, K. B., Effect of vortex bending and pinning on surface impedance of superconductors in the mixed state, *Physica B* **165&166** 1173-1174 (1990)

Geshkenbein V. B., Feigel'man M. V., Vinokur, V. M., Temporal decay and frequency response of the critical state in type II superconductors, *Physica C* **185-189** 2511-2512 (1991)

van Otterlo, A., Scalettar, R., and Zimányi, G., Phase diagram of disordered vortices from London Langevin simulations, *Phys. Rev. Lett.* **81** 1497 (1998).

Kierfeld, J., and Vinokur, V., Lindemann criterion and vortex lattice phase transitions in type-II superconductors, *Phys. Rev. B* **69** 024501 (2004).

Frey, E., Nelson, D., and Fisher, D., Interstitials, vacancies, and supersolid order in vortex crystals, *Phys. Rev. B* **49** 9723 (1994).

Ertas, D., and Nelson, D., Irreversibility, mechanical entanglement and thermal melting in superconducting vortex crystals with point impurities, Physica C **272** 79 (1996).

de Souza Silva, C. C., and Carneiro, G., Simple model for dynamical melting of moving vortex lattices interacting with periodic pinning, *Phys. Rev. B* **66** 054514 (2002)